\documentclass[a4paper,11pt]{article}
\pdfoutput=1 

\usepackage{jheppub} 

\usepackage[T1]{fontenc} 
\usepackage{physics}
\NewDocumentCommand{\tens}{t_}
{%
	\IfBooleanTF{#1}
	{\tensop}
	{\otimes}%
}
\NewDocumentCommand{\tensop}{m}
{%
	\mathbin{\mathop{\otimes}\displaylimits_{#1}}%
}
\usepackage{graphicx}
\usepackage{wrapfig}
\usepackage{amsmath}
\usepackage{amsfonts}
\usepackage{amssymb}
\usepackage{quotes}
\usepackage{subcaption}
\usepackage{xcolor}

\title{\textbf{\boldmath Information theoretic measures for Lifshitz system}}

\author{Souvik Paul,}
\author{Anirban Roy Chowdhury,}
\author[1]{Ashis Saha,\note{Corresponding author.}}
\author{Sunandan Gangopadhyay}

\affiliation[a]{\textit{Department of Astrophysics and High Energy Physics\\}
	\textit{ S.N.~Bose National Centre for Basic Sciences,\\}
	\textit{JD Block, Sector-III, Salt Lake, Kolkata 700106, India}}

\emailAdd{souvik.paul@bose.res.in}
\emailAdd{iamanirban@bose.res.in}
\emailAdd{ashis.saha@bose.res.in}
\emailAdd{sunandan.gangopadhyay@bose.res.in}

\abstract{\noindent In this work, we have studied various mixed state information theoretic quantities for an excited state of Lifshitz spacetime in $3+1$-dimensions. This geometry is the gravity dual to a class of $2+1$-dimensional quantum field theories having Lifshitz symmetry. We have holographically calculated mutual information, entanglement wedge cross section, entanglement negativity and mutual complexity for strip like subsystems at the boundary. For this we have used the results of holographic entanglement entropy and complexity present in the literature. We first calculate all of these mentioned quantities for the pure state of Lifshitz spacetime. Then we have moved on to calculate all these quantities for excited state of the Lifshitz spacetime. 
	The gravity dual of excited state of Lifshitz systems in field theory can be obtained by applying constant perturbations along the boundary direction. Further, we would like to mention that for the simplicity of calculation we are only considering results up to the first order in perturbation. The change in the obtained holographic information theoretic quantities are then related to entanglement entropy, entanglement pressure, entanglement chemical potential and charge using the stress tensor complex. These relations are analogous to the first law of entanglement thermodynamics given earlier in the literature. All the calculations are carried out for both values of dynamical scaling exponent ($z$) present in the Lifshitz field theory.}
\begin{document}
\maketitle
\flushbottom
\section{Introduction}
In recent years, the discovery of the $\mathrm{AdS/CFT}$ (anti-de Sitter/conformal field theory) correspondence \cite{Maldacena:1997re,Gubser:1998bc,Witten:1998qj,Aharony:1999ti,Natsuume:2014sfa,Nastase:2007kj} by Maldacena have provided us various tools to investigate strongly coupled quantum field theories using a classical gravity theory which is asymptotically an $\mathrm{AdS}$ spacetime in one special dimension higher. This correspondence provides deep insights in different fields like the field of black hole physics \cite{Emparan:2002px,Tanaka:2002rb,Gregory:2004vt}, quantum gravity \cite{Bak:2006nh,Rovelli:1997na,Silva:2023ieb,Engelhardt:2015gla}, quantum chromodynamics \cite{Erlich:2005qh,Karch:2006pv,Aharony:2002up,Kruczenski:2004me,Andreev:2006ct,Csaki:2008dt} and strongly coupled condensed matter systems   \cite{Hartnoll:2008vx,Hartnoll:2008kx,Herzog:2009xv,Horowitz:2010gk,Gregory:2009fj,Li:2011xja,Gangopadhyay:2012am,Gangopadhyay:2012np,Yang:2019gce}. For example, we can talk about $AdS_{5}/CFT_{4}$ correspondence, where $\mathcal{N}=4$ supersymmetric Yang–Mills theory in $(3+1)$-dimensions is dual to type IIB superstring theory on $AdS_{5}\cross S^{5}$. This correspondence has also guided us to compute various information theoretic quantities on the CFT side holographically. The understanding that information theoretic quantities can be computed holographically started with the work of Ryu and Takayangi \cite{Ryu:2006bv,Ryu:2006ef}. In this work, we shall take a look at the Lifshitz class of quantum field theories \cite{lifshitz1941theory,PhysRevB.23.4615,MohammadiMozaffar:2017nri,MohammadiMozaffar:2019gpn,MohammadiMozaffar:2018vmk,Vasli:2023syq} from a holographic perspective. These class of theories are important in condensed matter physics. The convincing reason behind studying Lifshitz system lies in the fact that, Lifshitz field theory plays a very important role across various contexts of physics starting from condensed matter physics to quantum gravity. Sometimes real world symmetries exhibit non-relativistic scale invariance near the critical point \cite{Coleman:2005hnw,Sachdev:2011cs,gegenwart2008quantum}, rather than relativistic invariance. As the Lifshitz field theory also possesses this kind of symmetry, it provides us a framework for understanding critical phenomenon and fixed points in quantum field theory. Some examples of non-relativistic type of quantum critical systems are heavy fermion compound YbRh$_2$Si$_2$ in spatial dimension $d=3$ and dynamical scaling exponent $z=4$ \cite{doi:10.1073/pnas.1200346109}, CeCu$_{5.9}$Au$_{0.1}$ for $d=2$ and $z=8/3$ \cite{PhysRevB.90.045105}. Also the Lifshitz scalar field theory and its generalizations have been extensively used to study quantum phase transitions in strongly correlated electron systems \cite{PhysRevB.72.024420,sachdev1999quantum,PhysRevLett.35.1678}. Researchers apply Lifshitz theory to understand emergent phenomena, such as superconductivity, magnetism, and other collective behaviors in condensed matter systems \cite{Chen:2009ka,Srivastav:2023qof,Brynjolfsson:2009ct,Lu:2013tza,Sin:2009wi,COUTINHOFILHO1980433}. On the other hand  Hořava theory of gravity \cite{Horava:2009uw,Sotiriou:2010wn,Li:2009bg} at Lifshitz point, tries to quantize gravity by using techniques of quantum field theory. This theory exhibits a dynamical critical exponent $z=3$ in the ultraviolet (UV) region. Lifshitz fixed points have also been used to study the physics of early universe and encounter questions related to flatness of universe \cite{Chen:2009ka}. Besides all these, the excited state of Lifshitz field theory have been extensively studied in \cite{Angel-Ramelli:2020xvd,Zhou:2016ekn,ARDONNE2004493}. These kind of systems also plays an important role in the study of quasi-particle excitation \cite{Castro-Alvaredo:2018bij,Castro-Alvaredo:2018dja,Castro-Alvaredo:2019lmj,Molter:2014hna}. People have also studied these systems with higher excitation. These systems demonstrate an unexpected lack of ergodicity and hold significance for the eigenstate thermalization hypothesis \cite{Bernien:2017ubn,Turner:2018kjz}. Our objective is to compute various information theoretic measures for the Lifshitz theories holographically. We therefore start by briefly reviewing some of the important quantities in the context of quantum information theory.

\noindent
Entanglement entropy (or von-Neumann entropy) is an important quantity in the context of the context of quantum information theory. The knowledge of entanglement entropy provides us enormous information about the underlying state of the quantum theory. To calculate the entanglement entropy in the context of quantum information theory, the procedure is as follows. One first chooses a pure bipartite quantum system with Hilbert space of the form $\mathcal{H}=\mathcal{H}_{A}\tens\mathcal{H}_{B}$, where $A\cup B$ is the full system. The entanglement entropy for a given state is then given by the von-Neumann entropy of the reduced density matrix, obtained by integrating over the degrees of freedom outside $A$. Hence, the von-Neumann entropy of the subsystem $A$ is given by \cite{von2013mathematische,nielsen2010quantum}
\begin{equation}
	S_{EE}(A) = -\Tr (\rho_A \log \rho_A)
\end{equation}
where $\rho_A$ is the reduced density matrix of the system $A$, which is obtained by tracing over the degrees of freedom of $B$. Symbolically, we can define this by $\rho_A = \mathrm{Tr}_{B} (\rho_{A B})$, $\rho_{A B}=\ket{\psi} \bra{\psi}$, where $\ket{\psi} \in \mathcal{H}$. 

\noindent
Keeping the above definition of entanglement entropy in mind, we now define another useful quantity in the context of quantum information theory, known as mutual information. The mutual information between two subsystems $A$ and $B$ can be defined in the following way \cite{nielsen2010quantum}
\begin{equation}
	I(A:B)=S_{EE}(A)+S_{EE}(B)-S_{EE}(A \cup B)
\end{equation}
where $S_{EE}(A)$, $S_{EE}(B)$ and $S_{EE}(A \cup B)$ denote the entanglement entropy of $A$, $B$ and $A \cup B$ respectively. It should be mentioned that mutual information is a UV finite quantity. It is to be mentioned that for a given operator $O_{A}$ in region $A$ and $O_{B}$ in region $B$, mutual information obey the following inequality \cite{Wolf:2007tdq}
\begin{equation}
	I(A:B)\geq \frac{\left(\left<O_{A}O_{B}\right>-\left<O_{A}\right>\left<O_{B}\right>\right)^2}{2||O_{A}||^{2}~||O_{B}||^{2}}~.
\end{equation}
Hence, mutual information is a measure of the collective correlations between two subsystems (both classical and quantum correlation). The picture for mixed states however is quite difficult.\\
For mixed states  von-Neumann entropy fails to capture the entanglement as it involves irrelevant classical correlations. Different correlation measures for mixed states has been suggested in the previous literature. Among them the entanglement of purification (EoP) is one of the most promising candidate. The purification process involves creating a pure state $\ket{\psi}$ from the mixed state density matrix $\rho_{AB}$ by introducing auxiliary degrees of freedom to the original Hilbert space $\mathcal{H}$. The EoP is defined as follows. 
Let $\rho_{AB}$ be the density matrix of a bipartite system $\mathcal{H}=\mathcal{H}_{A}\tens \mathcal{H}_{B}$. If $\ket{\psi}\in \mathcal{H}_{AA'BB'}=\mathcal{H}_{AA'}\tens \mathcal{H}_{BB'}$ is a purification of $\rho_{AB}$, that is, $\rho_{AB}=\Tr_{A'B'} \ket{\psi}\bra{\psi}$, then the EoP for $\rho_{AB}$ is defined as \cite{Terhal_2002}
\begin{equation}
	E_{P}(\rho)=E_{P}(A,B)=\underset{\ket{\psi}}{min}~S_{AA'}
\end{equation}
where minimization is over all possible states $\psi$, and $S_{AA'}$ is the von Neumann entropy of the reduced density matrix $\rho_{AA'}=\Tr_{BB'}\ket{\psi}\bra{\psi}$. For a pure state $\rho_{AB}$, no purification is needed, hence $E_{P}=S(A)=S(B)$. 
One of the important property of EOP is that it is bounded above by half of the mutual information \cite{Takayanagi:2017knl,Nguyen:2017yqw}
\begin{equation}
	E_{P}(A,B)\geq\frac{I(A:B)}{2}~.
\end{equation}
It has also been shown in \cite{Takayanagi:2017knl,Nguyen:2017yqw} that for a tripartite system $E_{P}$ also obeys the following inequalities
\begin{equation}
	E_{P}(A:B\cup C)\geq E_{P}(A:B)~
\end{equation}
\begin{equation}
	E_{P}(A,B\cup C)\geq \frac{I(A:B)}{2}+\frac{I(A:C)}{2}~.
\end{equation}
In quantum field theory it is very difficult to compute the EoP for mixed states. However, the holographic principle offers a geometric approach to determine this quantity. The holographic counterpart of EoP is known as  the entanglement wedge cross-section (EWCS) \cite{Takayanagi:2017knl,Nguyen:2017yqw,Jokela:2019ebz,BabaeiVelni:2019pkw}. This conjecture leads us to the "$E_{P}=E_{W}$"  duality. Nevertheless, it should be mentioned that a direct proof of this duality has not yet been found. 

\noindent
Another computable measure for mixed state correlation was proposed in \cite{Vidal:2002zz,Plenio:2005cwa,Zyczkowski:1998yd,Zyczkowski:1999iw}, and was termed as entanglement negativity. This quantity serves as a maximum limit on the amount of entanglement that can be distilled from a quantum system in a mixed state. Considering a tripartite system $A \cup B$, with $A=A_{1}\cup A_{2}$ , and $B=A^{c}$, the reduced density matrix for the subsystem $A$ can be obtained by tracing over the $B$ degrees of freedom, that is $\rho_A = \Tr_{B} (\rho_{AB})$, where $\rho_{AB}$ is the density matrix of the full system $A\cup B$. Computation of the entanglement negativity involves partial transpose of the density matrix with respect to one of the systems of the bipartite subsystem. The partial transpose with respect to $A_{2}$ is defined by
\begin{equation}
	\bra{e^{A_1}_{i}e^{A_2}_{j}}\rho^{T_{A_2}}_{A}\ket{e^{A_1}_{k}e^{A_2}_{l}}=\bra{e^{A_1}_{i}e^{A_2}_{l}}\rho_{A}\ket{e^{A_1}_{k}e^{A_2}_{j}}
\end{equation} 
where $\ket{e^{A_1}_{i}}$ and $\ket{e^{A_2}_{i}}$ are orthonormal basis vectors associated with the Hilbert spaces of $A_1$ and $A_2$ respectively. Entanglement negativity actually quantifies the extent to which $\rho^{T_{A_2}}_{A}$ is negative. This property actually signifies the term negativity. In quantum information theory, we can define two different quantities, one is negativity and another one is entanglement negativity or logarithmic negativity. The term negativity corresponds to the absolute value of the sum of negative eigenvalues of $\rho^{T_{A_2}}_{A}$, which is defined as follows \cite{Vidal:2002zz}
\begin{equation}
	N(\rho)=\frac{\Tr |\rho^{T_{A_2}}_{A}|-1}{2}~.
\end{equation} 
It can be shown that the above quantity vanishes for product states. On the other hand entanglement negativity between two disjoint intervals $A_1$ and $A_2$ is defined by \cite{Vidal:2002zz}
\begin{equation}
	E_{N}(\rho)=\ln \Tr |\rho^{T_{A_2}}_{A}|~.
\end{equation}
The entanglement negativity for a quantum field theory can be obtained by using replica technique given in \cite{Calabrese:2012ew}. The main goal of this technique is to determine $\Tr (\rho^{T_{A_2}}_{A})^{n_e}$ and take the replica limit (analytic continuation of even sequences of $n=n_e$ to $n_e \to 1$). This gives \cite{Calabrese:2012ew}
\begin{equation}
	E_{N}(\rho)=\underset{n_e \to 1}{lim}\ln \Tr (\rho^{T_{A_2}}_{A})^{n_e}~.
\end{equation}
This technique is only applicable in $(1+1)$-dimensional CFT. Hence for higher dimensional CFT(s) it is extremely difficult to compute $E_{N}$ using these replica techniques. The gauge/gravity duality gives us a convenient way to compute $E_N$ of CFT using its gravity dual. Some holographic conjectures to calculate entanglement negativity is discussed later in this paper. 

\noindent
Recently there has been a notable focus on quantum complexity \cite{Jefferson:2017sdb,Chapman:2017rqy,Hackl:2018ptj,Bhattacharyya:2018bbv,Khan:2018rzm}, as well as the various related mixed state information theoretic measures. In case of pure states, complexity is defined as the minimum number of gates chosen from pre-established set of gates which takes us from a certain reference state ($\ket{\psi}_{R}$) to a certain target state ($\ket{\psi}_{T}$). If $\ket{\psi}_{T}=\hat{U}\ket{\psi}_{R}$, then $\hat{U}$ represents the unitary operator that transforms the reference state to the target state. This $\hat{U}$ is constructed from a set of elementary gates. Hence, one can write the unitary operation as
\begin{equation}
	\hat{U}=g_{1}g_{2}g_{3}\dots g_{n-1}g_{n}
\end{equation}  
where $g_{i}$ denotes an elementary quantum gate. The circuit complexity of the target state, represented by $C(\ket{\psi}_{T})$, is defined as the smallest number of gates required to create the unitary operation. This is one way to compute quantum complexity, although there are other techniques to calculate quantum complexity as mentioned in \cite{nielsen2005geometric,doi:10.1126/science.1121541,Ali:2018fcz,Susskind:2018pmk}. In this paper, we have discussed different holographic methods to compute complexity. 

\noindent
The concept of purification can also be introduced in the context of complexity. The purification complexity is defined by the minimal pure state complexity among all possible purifications available for the mixed state. For a bipartite mixed state, the purification complexity can be obtained by computing mutual complexity between two subsystems. For a bipartite system ($AB$), the mutual complexity is defined as\cite{Alishahiha:2018lfv,Caceres:2018blh,Caceres:2019pgf,Chen:2018mcc,Agon:2018zso}
\begin{equation}
	\Delta C= C(\rho_{A})+C(\rho_{B})-C(\rho_{A \cup B})~
\end{equation}
where $C(\rho_{i})$ is the complexity of the ith subsystem.
The mutual complexity is said to be subadditive if $\Delta C>0$ and superadditive if $\Delta C<0$ .

\noindent
In this paper, we have computed the various mixed state information theoretic quantities discussed above for an excited state of the Lifshitz field theory using the gauge/gravity duality. We find that the results of different information theoretic quantities get modified for excited Lifshitz theory. We have also graphically presented all these results. 
In a system which is away from equilibrium, it is natural to ask a question whether there exists a first law like relation of thermodynamics between these quantities. Various studies in this direction\cite{Allahbakhshi:2013rda,Bhattacharya:2012mi,Mukohyama:1997ww,Wong:2013gua}, have addressed this question. It was shown that for a sufficiently small subsystem, change in entanglement entropy is proportional to change in energy density. This constant of proportionality can be identified as some sort of temperature which is dependent on the width of the subsystem. This quantity is named as "$entanglement~ temparature$" in \cite{Bhattacharya:2012mi,Mukohyama:1997ww}. This kind of simple first law like relation of entanglement thermodynamics only holds for systems with rotational and translational invariance\cite{Bhattacharya:2012mi}. If the system is not rotationally invariant, the entanglement entropy is not only proportional to change in energy density but also depends on the change in entanglement pressure and the entanglement chemical potential. This kind of problem was addressed in \cite{Allahbakhshi:2013rda,Guo:2013aca,Jeong:2022jmp,Jeong:2022zea}.

\noindent
All of these studies have developed entanglement thermodynamic relations for relativistic theories. In \cite{Chakraborty:2014lfa,Karar:2017org}, the authors have chosen an exited state of a non-relativistic system (Lifshitz system in $3+1$-dimensions), and they have related the change in holographic entanglement entropy and holographic complexity with change in energy, pressure and chemical potential. These relations are equivalent to first law of entanglement thermodynamics for this scenario. In this paper, we have established first law like relationships of entanglement thermodynamics for the change in EWCS, entanglement negativity, and mutual complexity in an excited state of Lifshitz system.  
\section{Lifshitz spacetime geometry}\label{section 2}
The pure $AdS$ spacetime is a solution of the vacuum Einstein field equation of general relativity with negative cosmological constant. On the other hand to obtain Lifshitz spacetime geometry one needs a non-vanishing energy momentum tensor, that is, one needs to add matter field in Einstein's field equation. There are various different choices of the matter fields\cite{Taylor:2008tg,Kachru:2008yh,Korovin:2013bua}. Among the various different choices of matter fields, the massive vector field is the one which is interesting in its own right. In this set up, the Einstein-Hilbert action can be written as \cite{Kachru:2008yh,Taylor:2008tg}
\begin{equation}\label{action unpurturbed lifshitz}
	S=\frac{1}{16\pi G_{4}}\int d^4x \sqrt{-g}\left(R-2\Lambda-\frac{1}{4}F_{\mu\nu}F^{\mu\nu}-\frac{1}{2}m^2 A_{\mu}A^{\mu}\right)~
\end{equation}
where $R$ is the Ricci scalar, $\Lambda$ is the cosmological constant, $F_{\mu\nu}$ is the Maxwell field strength tensor, and $A_{\mu}$ is the massive gauge field. Now varying the above action with respect to the metric $g_{\mu\nu}$, one gets the following equations of motion \cite{Ross:2011gu}
\begin{equation}\label{eom gravity}
	R_{\mu\nu}=\Lambda g_{\mu\nu}+\frac{1}{2}F_{\mu\sigma}F_{\nu}^{\sigma}-\frac{1}{8}F_{\sigma\rho}F^{\sigma\rho}g_{\mu\nu}+\frac{1}{2}m^2 A_{\mu}A_{\nu}~.
\end{equation}
Similarly varying the action with respect to the massive gauge field ($A_{\mu}$), one gets
\begin{equation}\label{eom gauge field}
	\grad_{\mu}{F^{\mu\nu}}=m^2 A^{\nu}~.
\end{equation} 
If we choose $\Lambda=-\frac{1}{2}(z^2+z+4)$ and $m^2=2z$ \cite{Kachru:2008yh,Taylor:2008tg}, the solution of the equation of motions in eq.(s)(\eqref{eom gravity},\eqref{eom gauge field}) gives the following line element 
\begin{align}
	ds^2 &= -r^{2z}dt^2 + r^2(dx^2+dy^2)+\frac{dr^2}{r^2}~\label{pure lifshitz metric},\\
	&A=\alpha r^{z}dt,~~\alpha^2=\frac{2(z-1)}{z}\label{ground metric}~.
\end{align}	
This solution is known as the Lifshitz metric in $3+1$-spacetime dimensions. The spacetime geometry is the gravitational dual to $2+1$-dimensional field theory with Lifshitz symmetry near the quantum critical point. The Lifshitz scaling symmetry of the solution reads
\begin{equation}
	t\to\lambda^{z}t,~ x\to\lambda x,~ y\to\lambda y,~ r\to \lambda^{-1}r
\end{equation}
where $z$ is the dynamical scaling exponent. The symmetry is denoted by $Lif^{z}_{3+1}$. 
\subsection{Holographic computation of entanglement entropy}\label{HEE 2.1}
\noindent
In this subsection, we present a brief review of the computation of the HEE for the Lifshitz spacetime.
Let us consider a conformal field theory in $d$-dimension, hence the gravity dual is $(d+1)$-dimensional. The gravity dual of the entanglement entropy of a CFT living at the boundary is nothing but the holographic entanglement entropy (HEE). Now for computing the HEE, we will follow the Ryu-Takayanagi (RT) prescription \cite{Ryu:2006bv,Ryu:2006ef}.
The RT formula relates the entanglement entropy of a strongly coupled CFT with the area of a codimension-two static minimal surface in the bulk. This co-dimension two static minimal surface is known as the RT surface.\\
The holographic entanglement entropy for $A$ is given by \cite{Ryu:2006bv,Ryu:2006ef}\footnote{The RT formula for the HEE is applicable to the Lifshitz geometry for the static case \cite{Solodukhin:2009sk,Nesterov:2010yi}. The reason for this is that the AdS/CFT duality works for any spacetime geometry which is asymptotically AdS.} 
\begin{equation}
	S_{HEE}(A)=\frac{1}{4G^{(d+1)}_{N}}[Area(\Gamma^{A}_{min})]
\end{equation}
where $G^{(d+1)}_{N}$ is the Newton's gravitational constant in $(d+1)$-spacetime dimensions and $\Gamma^{A}_{min}$ is the $(d-1)$-dimensional static minimal surface such that $\partial \Gamma^{A}_{min}=\partial A$.\\ 
We will now proceed to calculate the HEE for a striplike subsystem  $A$. The entangling region at the boundary is taken to be a strip of length $l$ ($-\frac{l}{2}\leq x\leq\frac{l}{2}$) and width $L$ ($0\leq y\leq L$). This means we will vary the length along the $x$-direction by keeping the width of the subsystem to be fixed. To calculate the RT area functional, we will first choose a constant time slice. This means that we will set $dt=0$. Then we will parameterize the bulk coordinate in terms of the boundary, that is, $r=r(x)$. Under this consideration, the RT area functional for the metric given in eq.\eqref{pure lifshitz metric} reads
\begin{equation}
	A^{(0)}=2L\int_{0}^{l/2} dx~r^2 \sqrt{1+\frac{r'(x)^2}{r^4}}\label{area functional ground}~.
\end{equation}
We can identify the integrand of the above area functional to be the Lagrangian of the form $\mathcal{L}=\mathcal{L}(r,r^{'})$. The above expression of the integrand suggests that $x$ is a cyclic coordinate.  This leads to the following conserved quantity
\begin{equation}
	H=\frac{-r^{2}}{\sqrt{1+\frac{r^{'}(x)^2}{r^4}}}=constant=c~.
\end{equation}
Now at the turning point ($r^{(0)}_{t}$) in the bulk $r^{'}(x)=0$. This fixes the above constant to be $c=-r^{(0)2}_{t}$. This results in the following equation for $r^{'}(x)$
\begin{equation}
	r'(x)=\frac{r^4}{r_{t}^{(0)2}} \sqrt{1-\frac{r_{t}^{(0)4}}{r^4}}\label{r prime ground}~,~r'(x)=\frac{dr}{dx}~.
\end{equation}
The profile of the extremal surface $x(r)$ can be obtained by integrating the above equation
\begin{equation}
	x(r)=\int_{r_{t}^{(0)}}^{r} du \frac{r_{t}^{(0)2}}{u^4}\frac{1}{\sqrt{1-\frac{r_{t}^{(0)4}}{u^4}}}~.\label{profile}
\end{equation}
Substituting eq.\eqref{r prime ground} in eq.\eqref{area functional ground}, we get
\begin{align}
	A^{(0)}&=2L\int_{r_{t}^{(0)}}^{\delta} dr \frac{\frac{r^2}{r_{t}^{(0)2}}}{\sqrt{\frac{r^4}{r_{t}^{(0)4}}-1}}\nonumber\\
	&\approx 2L \delta - \frac{5}{3} L r_{t}^{(0)}\label{A0 25}~
\end{align}
where $\delta$ is the UV cutoff.
In the above expression we have ignored terms of $O(1/\delta)$. Hence from the RT proposal, the HEE is given by\cite{Chakraborty:2014lfa}
\begin{equation}
	S^{(0)}_{HEE}=\frac{A^{(0)}}{4G_{4}}=\frac{L}{4G_{4}}\left(2\delta -\frac{5}{3}r^{(0)}_{t}\right)\label{HEE pure}~.
\end{equation} 

\noindent
The subsystem length ($l$) in terms of the bulk coordinate can be calculated from eq.\eqref{profile}. The subsystem length in terms of the turning point reads
\begin{equation}
	\frac{l}{2}=\int_{r_{t}^{(0)}}^{\infty} dr \frac{\frac{r_{t}^{(0)2}}{r^4}}{\sqrt{1-\frac{r_{t}^{(0)4}}{r^4}}}=\frac{\sqrt{\pi}\Gamma(3/4)}{\Gamma(1/4)r_{t}^{(0)}}=\frac{1}{N r_{t}^{(0)}}\label{length pure}~
\end{equation}
where $N=\left(\int_{1}^{\infty}\frac{d\eta}{\eta^2 \sqrt{\eta^4-1}}\right)^{-1}=\frac{\Gamma(1/4)}{\sqrt{\pi}\Gamma(3/4)}$ is a constant which will be useful throughout the calculation.
We can represent the HEE in terms of the subsystem length by using eq.(s)(\eqref{HEE pure},\eqref{length pure}). The expression of $S_{HEE}$ in terms of the subsystem length reads 
\begin{equation}\label{HEE0 pure}
	S^{(0)}_{HEE}(l)=\frac{L}{4G_{4}}\left(2\delta -\frac{10}{3Nl}\right)~.
\end{equation} 

\noindent
\subsection{Holographic mutual information and EWCS}\label{12}
The holographic calculation of mutual information (HMI) can be carried out as follows. The HMI between two subsystems A and B reads
\begin{equation}
	I(A:B)=S_{HEE}(A)+S_{HEE}(B)-S_{HEE}(A \cup B)\label{mutual information formula}~
\end{equation}
where $S_{HEE}(A)$, $S_{HEE}(B)$ and $S_{HEE}(A\cup B)$ are HEE of subsystems $A$, $B$ and $A\cup B$ respectively.
For two strip like subsystems (A and B) of width $l$ and separated by a distance $d$ (such that $d/l<1$), the expression for mutual information is given by
\begin{equation}
	I(A:B)=2S_{HEE}(l)-S_{HEE}(2l+d)-S_{HEE}(d)~\label{mutual ground}
\end{equation}
where we have used $S_{HEE}(A \cup B)=S_{HEE}(2l+d)+S_{HEE}(d)$.
\begin{figure}[!h]
	\centering
	\includegraphics[width=12cm]{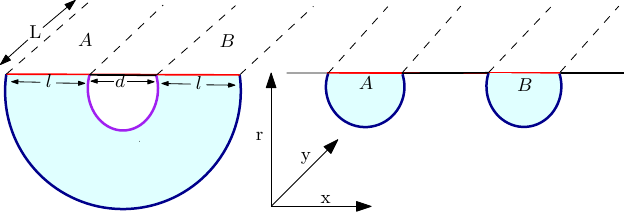}
	\caption{A schematic diagram of possible RT hyper surfaces for two strip like parallel subsystems in the boundary. The diagram on the left denotes connected phase and the diagram on the right denotes disconnected phase.}
	\label{fig:connected and disconnected phase}
\end{figure}
\noindent
For unperturbed Lifshitz spacetime, we can compute HMI using eq.\eqref{mutual ground}, the expression reads\cite{Fischler:2012uv}
\begin{align}
	I^{(0)}(A:B)&=\frac{L}{4G_{4}}\left[2\left(2\delta-\frac{10}{3Nl}\right)-\left(2\delta-\frac{10}{3N(2l+d)}\right)-\left(2\delta-\frac{10}{3Nd}\right)\right]\\
	&=\frac{L}{4G_{4}}\frac{10}{3N}\left[\frac{1}{2l+d}+\frac{1}{d}-\frac{2}{l}\right]~.
\end{align}
For simplicity of calculation, we will define $\bar{I}^{(0)}(A:B)=\frac{4G_{4}}{L}I^{(0)}(A:B)$, hence we have
\begin{equation}
	\bar{I}^{(0)}(A:B)=\frac{10}{3N}\left[\frac{1}{2l+d}+\frac{1}{d}-\frac{2}{l}\right]~.
\end{equation}

\noindent
Now we will begin our discussion on the construction of entanglement wedge cross section in AdS/CFT as mentioned in \cite{Takayanagi:2017knl,Nguyen:2017yqw,Jokela:2019ebz,BabaeiVelni:2019pkw}. For simplicity we will assume two strip like subsystems (A and B) of equal width $l$ and separated by a distance $d$ on the boundary $\delta M$ ($\delta M$ is the boundary of the canonical time slice $M$ which is considered to be the gravity dual). Let $\Gamma^{min}_{A}$, $\Gamma^{min}_{B}$ and $\Gamma^{min}_{AB}$ are the RT surfaces corresponding to $A$, $B$ and $A \cup B $ respectively. We will also assume that there is no overlap between the two subsystems, that is, $A \cap B=0$
\cite{Jokela:2019ebz}. The bulk region surrounded by $AB$ and $\Gamma^{min}_{AB}$ is called the entanglement wedge $M_{AB}$. The boundary of entanglement wedge $M_{AB}$ is characterized by 
\begin{equation}
	\delta M_{AB}=A\cup B \cup \Gamma^{min}_{AB}=\bar{\Gamma}_{A} \cup \bar{\Gamma}_{B}~.
\end{equation}
It should be mentioned that if $A$ and $B$ are small enough or the separation between them is large enough, the entanglement wedge $M_{AB}$ becomes disconnected. Entanglement wedge cross section (EWCS) is defined by area of the minimal surface separating $M_{AB}$ into two bulk subregions associated with $A$ and $B$ respectively. Now we will rewrite $\Gamma^{min}_{AB}$ as follows
\begin{equation}
	\Gamma^{min}_{AB}=\bar{\Gamma}_{A} \cup \bar{\Gamma}_{B}~.
\end{equation}
Here we have used $\bar{\Gamma}_{A}=A \cup \Gamma^{A}_{AB} $, $\bar{\Gamma}_{B}=B \cup \Gamma^{B}_{AB}$ and $\Gamma^{min}_{AB}=\Gamma^{A}_{AB} \cup \Gamma^{B}_{AB}$.
Using the above equation, the boundary of the entanglement wedge can be written as 
\begin{equation}
	\delta M_{AB}=\bar{\Gamma}_{A} \cup \bar{\Gamma}_{B}~.
\end{equation}
Finally, the minimal area surface $\Sigma_{AB}^{min}$ which terminates at the boundary of the entanglement wedge, dividing the entanglement wedge in two parts with the following condition
\begin{equation}
	\delta \Sigma_{AB}^{min}=\delta \bar{\Gamma}_{A} = \delta \bar{\Gamma}_{B} 
\end{equation}
is called the entanglement wedge cross section. 
\begin{figure}[!h]
	\centering
	\includegraphics[width=8cm]{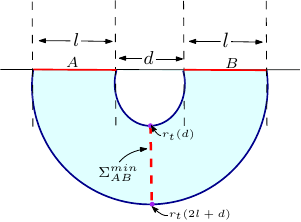}
	\caption{A schematic diagram of two thin strip like boundary subsystems of equal width $l$ and separated by a distance $d$, the shaded region denotes the entanglement wedge and the red dashed line indicates entanglement wedge cross section. The two points at $r_{t}(2l+d)$ and $r_{t}(d)$ denotes turning points of the RT surfaces corresponding to $\Gamma^{min}_{(l\cup l)}$ and $\Gamma^{min}_{d}$ respectively. }
	\label{fig:ewcs}
\end{figure}
\noindent
The entanglement wedge can be splitted into two subregions in infinite possible ways and thus there is a infinite set of possible $\Sigma_{AB}^{min}$. The EWCS is calculated by minimizing the area of $\Sigma_{AB}^{min}$ over all possible choices for $\Sigma_{AB}^{min}$. Then the expression for EWCS ($E_{W}$) reads \cite{Takayanagi:2017knl}
\begin{equation}
	E_{W}(A:B)=\underset{\bar{\Gamma}_{A}\subset \delta M_{AB}}{min}\left[\frac{A(\Sigma_{AB}^{min})}{4G}\right]\label{EW general formula}~.
\end{equation} 
Some of the properties of EWCS can be found in \cite{Takayanagi:2017knl,Nguyen:2017yqw,KumarBasak:2020eia}.
Now using eq.\eqref{EW general formula}, we will calculate EWCS for ground state of Lifshitz space time in $3+1$- dimensions. To do so we have to evaluate vertical constant $x$ hypersurface with minimal area which splits $M_{AB}$ into two sub-regions corresponding to $A$ and $B$. The induced metric on the constant $x$ hypersurface is given by
\begin{equation}
	ds_{ind}^2=r^2 dy^2 + \frac{dr^2}{r^2}\label{ind ewcs}~.
\end{equation}
By using the above induced metric and the eq.\eqref{EW general formula}, the EWCS is found to be
\begin{align}
	E^{(0)}_{W}&=\frac{1}{4G_4}\int_{0}^{L} dy \int_{r_{t}^{(0)}(2l+d)}^{r_{t}^{(0)}(d)} dr \nonumber\\ 
	&=\frac{L}{4G_4}\left[r_{t}^{(0)}(d)-r_{t}^{(0)}(2l+d)\right]\nonumber~.
\end{align}
Now using eq.\eqref{length pure}, we can represent the above result in terms of subsystem length as follows
\begin{equation}
	E^{(0)}_{W}=\frac{L}{4G_4}\left(\frac{2}{N}\right)\left[\frac{1}{d}-\frac{1}{2l+d}\right]~.
\end{equation}
Now we will recast the above result of EWCS in the following way
\begin{equation}
	\bar{E}^{(0)}_{W}=\left(\frac{2}{N}\right)\left[\frac{1}{d}-\frac{1}{2l+d}\right]~
\end{equation}
where $\bar{E}^{(0)}_{W}=\frac{4G_{4}}{L}E^{(0)}_{W}$.
\subsection{Holographic computation of Entanglement Negativity}\label{2.3}
In this subsection, we shall discuss about the procedure of computing entanglement negativity holographically. In this regard various literatures have suggested two different proposals.

\noindent
In one of the proposals the authors of \cite{Kudler-Flam:2018qjo,Kusuki:2019zsp} suggested that entanglement negativity($E_N$) is dual to EWCS backreacted by an extremal cosmic brane that terminates at the boundary of the entanglement wedge. This proposal is motivated by quantum error correcting codes. Hence the logarithmic negativity is equivalent to cross-sectional area of the entanglement wedge plus a bulk correction term. Although for a general entangling surface, calculation of negativity is very difficult due to the backreaction of the cosmic brane. The calculation of $E_N$ is simpler for spherical entangling surface. The holographic entanglement negativity corresponding to the dual CFT reads \cite{Kusuki:2019zsp,Kudler-Flam:2018qjo}
\begin{equation}
	E_N=\chi_{d}\frac{E_W}{4G_N}+E_{bulk}
\end{equation}
where $\chi_{d}$ is a dimension dependent constant\footnote{For subsystems with spherical entangling surface, an explicit mathematical expression for $\chi_{d}$ can be found in \cite{Hung:2011nu}.}.

\noindent
In this paper we will follow another proposal that suggests entanglement negativity is nothing but certain combinations of co-dimension two RT surfaces\cite{Jain:2017uhe,Jain:2017xsu,Chaturvedi:2016rcn,Chaturvedi:2016rft}, this type of combination can be obtained from the dual CFT correlators. It is quite interesting that both of these proposals leads to the exact same result of entanglement negativity in CFT \cite{Rogerson:2022yim,Matsumura:2022ide,Bertini:2022fnr,Roik:2022gbb,Dong:2021oad,Afrasiar:2021hld,Bhattacharya:2021dnd,Hejazi:2021yhz}

\noindent
Now we will calculate entanglement negativity holographically for two different scenarios. At first we will calculate $E_N$ for two adjacent subsystems then we will calculate the same for disjoint subsystems.

\noindent
Let us consider two parallel strip like adjacent subsystems A and B of width $l_1$ and $l_2$ respectively. The entanglement negativity for this system is given by \cite{Jain:2017uhe,Jain:2017xsu,Chaturvedi:2016rcn,Chaturvedi:2016rft}
\begin{equation}
	E_{N_{adj}}=\frac{3}{4} \left[S_{HEE}(l_1)+S_{HEE}(l_2)-S_{HEE}(l_1+l_2)\right]\label{negetivity attached}~
\end{equation} 
where $S_{HEE}(l_1)$, $S_{HEE}(l_2)$ and $S_{HEE}(l_1 + l_2)$ are the HEE of subsystems of length $l_1$, $l_2$ and $l_1 + l_2$ respectively. 
Hence, the entanglement negativity for two strip like adjacent boundary subsystems in case of unperturbed Lifshitz system reads
\begin{equation}
	E^{(0)}_{N_{adj}}=\frac{3L}{16G_{4}} \left[2\delta- \frac{10}{3N}\left(\frac{1}{l_1}+\frac{1}{l_2}-\frac{1}{l_1+l_2}\right)\right]~.
\end{equation}
For two subsystems with equal length, that is, $l_1 =l_2 =l$, the above equation becomes
\begin{equation}
	E^{(0)}_{N_{adj}}=\frac{3L}{16G_{4}}\left[2\delta -\frac{5}{Nl}\right]~.
\end{equation}
It can be seen that entanglement negativity for two adjacent subsystem contains a divergent piece. 

\noindent
Now we will proceed to compute entanglement negativity for two disjoint subsystems holographically. To do so, we will consider two strip like subsystems $A$ and $B$ of length $l_1$ and $l_2$ respectively and separated by a distance $d$. In this setup the expression of entanglement negativity reads \cite{KumarBasak:2020viv,Malvimat:2018ood,Afrasiar:2021hld}
\begin{equation}
	E_{N_{dis}}=\frac{3}{4} \left[S_{HEE}(l_1+d)+S_{HEE}(l_2+d)-S_{HEE}(l_1+l_2+d)-S_{HEE}(d)\right]\label{negetivity disjoint}~.
\end{equation}
Now using eq.\eqref{HEE0 pure} in the above equation, we can compute the  entanglement negativity for two disjoint subsystems. Hence the expression of entanglement negativity for disjoint subsystems reads
\begin{equation}
	E^{(0)}_{N_{dis}}=\frac{3L}{16G_{4}}\left[\frac{10}{3N}\left(\frac{1}{l_1 +l_2 +d}+\frac{1}{d}-\frac{1}{l_1  +d}-\frac{1}{l_2 +d}\right)\right]~.
\end{equation} 
If we consider that, the subsystem having equal length, that is, $l_1$ = $l_2$ = $l$, then the entanglement negativity is given by
\begin{equation}
	\bar{E}^{(0)}_{N_{dis}}=\frac{3}{4}\left[\frac{10}{3N}\left(\frac{1}{2l +d}+\frac{1}{d}-\frac{2}{l  +d}\right)\right]
\end{equation}
where we have used $\bar{E}^{(0)}_{N_{dis}}=\frac{4G_{4}}{L}E^{(0)}_{N_{dis}}$. 
Unlike the adjacent scenario, the entanglement negativity for two disjoint subsystems is independent of the  cutoff $\delta$.
\subsection{Holographic subregion complexity(HSC)}\label{2.4}
In our previous discussions, we have briefly revised the concept of quantum circuit complexity in the context of quantum information theory. In this subsection, we will discuss various holographic proposals aimed to calculate quantum complexity of boundary states. 

\noindent
In \cite{Susskind:2014moa,Susskind:2014rva,Susskind:2018pmk} it was proposed that the complexity of the boundary states is proportional to the volume of Einstein Rosen bridge (ERB) connecting the boundaries of a two sided eternal black hole. According to this proposal, we get the following relation
\begin{equation}
	C_{V}(t_L,t_R)=\frac{V^{ERB}(t_L,t_R)}{8 \pi R G_{d+1}}
\end{equation}
where $R$ is the $AdS$ radius and $V^{ERB}(t_L,t_R)$ is the co-dimension one extremal volume of ERB which is
bounded by the two spatial slices at times $t_L$ and $t_R$ of two CFTs that live on the boundaries of the eaternal black hole. The above equation is also called as $"Complexity=Volume"$ conjecture.

\noindent 
Another proposal \cite{Brown:2015bva,Brown:2015lvg,Goto:2018iay,Alishahiha:2018lfv}
states that complexity can be obtained from the bulk action
evaluated on the Wheeler-DeWitt patch which is bounded by the light sheets. This proposal is named as $"Complexity=Action"$ conjecture. This conjecture gives the following relation
\begin{equation}
	C_A=\frac{\mathcal{I}_{WdW}}{\pi \hbar}~.
\end{equation}
In our analysis we will use the conjecture proposed in \cite{Alishahiha:2015rta,Stanford:2014jda}, this tells holographic subregion complexity is proportional to the volume enclosed between RT surface and the boundary. If $V(\gamma)$ is the volume enclosed by minimal hypersurface in the bulk, then the holographic complexity becomes
\begin{equation}
	C_{A}=\frac{V(\gamma)}{8 \pi R G_{d+1}}~.
\end{equation} 
We will now proceed to compute HSC for ground state of Lifshitz spacetime.
As mentioned in eq.\eqref{profile} the profile of the  extremal surface is described by $x=x(r)$. The volume enclosed by the RT surface in the bulk is given by
\begin{equation}
	V^{(0)}=2L\int_{r^{(0)}_{t}}^{\infty}dr~r~x(r)~.
\end{equation} 
Substituting the expression of $x(r)$ from eq.\eqref{profile} in the above expression and introducing a cutoff $\delta$ for $r$ integral, we get\cite{Karar:2017org}
\begin{align}
	V^{(0)}&=2L\int_{r^{(0)}_{t}}^{\delta} dr~r\int_{r^{(0)}_{t}}^{r} du \frac{r_{t}^{(0)2}}{u^4}\frac{1}{\sqrt{1-\frac{r_{t}^{(0)4}}{u^4}}}\\
	&=\frac{\sqrt{\pi}\Gamma(\frac{3}{4})L \delta^2}{\Gamma(\frac{1}{4})r^{(0)}_{t}}-\frac{\sqrt{\pi}\Gamma(\frac{5}{4}) L r^{(0)}_{t}}{\Gamma(\frac{3}{4})}~.
\end{align}
In the above expression terms of $O(1/\delta)$ are ignored. It is known that, holographic complexity (HC) is proportional to the volume enclosed by the RT surface, hence the expression for HC reads \cite{Karar:2017org}
\begin{equation}
	C^{(0)}_{A}=\frac{\sqrt{\pi}}{8 \pi G_4 }\left[\frac{\Gamma(\frac{3}{4})L \delta^2}{\Gamma(\frac{1}{4})r^{(0)}_{t}}-\frac{\Gamma(\frac{5}{4}) L r^{(0)}_{t}}{\Gamma(\frac{3}{4})}\right]\label{complexity unperturbed}~.
\end{equation}
The above expression of $C^{(0)}_{A}$ can be represented in terms of the subsystem length ($l$) using eq.\eqref{length pure}, and that reads
\begin{equation}
	C^{(0)}_{A}=\frac{\sqrt{\pi}L}{8 \pi G_4 }\left[\frac{Nl\Gamma(\frac{3}{4}) \delta^2}{2\Gamma(\frac{1}{4})}-\frac{2\Gamma(\frac{5}{4})  }{Nl\Gamma(\frac{3}{4})}\right]~. \label{complexity unperturbed in l}
\end{equation}
\subsection{Mutual complexity}\label{2.5}
In this subsection, we shall proceed to calculate the mutual complexity.
In order to compute the complexity for mixed states we will follow the purification complexity protocol. Purification complexity is defined as the minimal complexity among all possible purifications of the mixed state. It suggests the optimization of the circuits, which takes the reference state to target space $\ket{\psi_{AB}}$ (which is a purification of the mixed state $\rho_{A}$) and also optimization over all possible purifications of $\rho_{A}$.
Now starting with a pure state $\rho_{AB}$ on an extended Hilbert space (original Hilbert space with auxiliary degrees of freedom), then tracing over the $B$ degrees of freedom, one can obtain the mixed state $\rho_A$. The purification complexity is expressed as \cite{Agon:2018zso}
\begin{equation}
	C(\rho_{A})=min_{B}C(\ket{\psi_{AB}})
\end{equation}
where $\rho_{A}=\Tr_{B}\ket{\psi_{AB}}\bra{\psi_{AB}}$. Mutual complexity is defined in order to compute the above mentioned mixed state complexity. 
In \cite{Alishahiha:2018lfv,Caceres:2018blh,Caceres:2019pgf,Chen:2018mcc,Agon:2018zso}, it is shown that, the purification complexity can be obtained by computing the mutual complexity between two subsystems. Therefore, the mutual complexity between two subsystems is defined as \cite{Alishahiha:2018lfv}
\begin{equation}
	\Delta C=C(\rho_A)+C(\rho_B)-C(\rho_{A \cup B})~.
\end{equation}
The mutual complexity is denoted to be superadditive if $\Delta C<0$ and subadditive if  $\Delta C>0$.\\
We will now proceed to calculate the mutual complexity for ground state of the Lifshitz spacetime.
To do so we first consider two subsystems $A$ and $B$ of equal width $l$, and which are separated by a distance $d$ on the boundary Cauchy slice. For $d/l <<1$, the mutual complexity is given by \cite{Agon:2018zso,Ben-Ami:2016qex,Ghodrati:2019hnn} 
\begin{equation}
	\Delta C=2C(l)-C(2l+d)+C(d)\label{mutual complexity}~
\end{equation}
where $C(l)$, $C(2l+d)$ and $C(d)$ are the complexity of subsystems of length $l$, $2l+d$ and $d$ respectively.
We would like to mention that, in order to compute $C(\rho_{A \cup B})$, we have used the fact that in the limit $\frac{d}{l}<<1$, $C(\rho_{A \cup B})$ can be written as \cite{Agon:2018zso,Ben-Ami:2016qex,Ghodrati:2019hnn}
\begin{equation}
	C(\rho_{A \cup B})=C(2l+d)-C(d)~.
\end{equation}

\noindent
Now we will proceed to compute mutual complexity for two strip like subsystems of equal size ($l$), and which are separated by a distance $d$ for unperturbed Lifshitz spacetime. The expression of mutual complexity can be calculated using eq.\eqref{mutual complexity} and eq.\eqref{complexity unperturbed in l}, and the expression reads\cite{Saha:2021kwq,Alishahiha:2018tep}
\begin{equation}
	\Delta C^{(0)}=\frac{\sqrt{\pi}L}{4\pi G_{4}}\frac{\Gamma(\frac{5}{4})}{N\Gamma(\frac{3}{4})}\left[\frac{1}{2l+d}-\frac{2}{l}-\frac{1}{d}\right]~.\label{mutual complexity unperturbed}
\end{equation}
The above result suggests that, the mutual complexity is independent of the UV cutoff. 

\section{Excited Lifshitz spacetime}\label{section 3}
In the previous section we have computed different information theoretic quantities for the pure Lifshitz spacetime. We found that none of these quantities depend on the dynamical exponent. The effect of the dynamical exponent on different information theoretic quantities can be observed if one considers the excited state of Lifshitz spacetime as suggested in \cite{Ross:2009ar,Ross:2011gu}.
It is to be noted that this effect also appears in the Lifshitz black brane \cite{Gong:2020pse}, where the dynamical exponent enters through the lapse function of the black brane geometry.
We will start this section by a brief review of the excited Lifshitz theory and the corresponding energy-momentum tensors derived from the boundary term of the action.

\noindent
In order to have a well-defined variational problem, it is necessary to include a boundary term in the action. The complete action can be written as follows:
\begin{align}
	S_{tot}=\frac{1}{16\pi G_{4}}\int d^4x \sqrt{-g}\left(R-2\Lambda-\frac{1}{4}F_{\mu\nu}F^{\mu\nu}-\frac{1}{2}m^2 A_{\mu}A^{\mu}\right)\nonumber\\
	+\frac{1}{16\pi G}\int d^3 \zeta\left(2K-4-z\alpha\sqrt{-A_{\alpha}A^{\alpha}}\right)+S_{deriv}\label{total action}
\end{align}
where $\zeta^{\alpha}$ are boundary coordinates at constant $r$, $h_{\alpha \beta}$ is the induced metric on the boundary, $K_{\alpha \beta }=\grad _{(\alpha}n_{\beta )}$ is the extrinsic curvature of the boundary with $n^{\alpha}$ being an orthogonal unit vector to the boundary in the outward direction. $K$ is the trace of $K_{\alpha \beta }$ with respect to the induced metric $h_{\alpha\beta}$, such that $K=h^{\alpha \beta}K_{\alpha \beta }$. $S_{deriv}$ consists of terms that include derivatives of the boundary fields, which may consist both the curvature tensor derived from the boundary metric and covariant derivatives of $A_{\alpha}$. Since the boundary is flat, the boundary fields remain constant, implying that $S_{deriv}$ does not impact the on-shell action or its first variation around the Lifshitz background. 

\noindent
The boundary term in the above action does not contribute to the equation of motion but it is useful to derive an expression for the stress tensor. In asymptotic Lifshitz spacetimes, the boundary theory exhibits non-relativistic behaviour, allowing for the definition of a stress tensor complex instead of a fully covariant stress tensor. This stress tensor complex includes the energy density, energy flux, momentum density, and spatial stress tensor. 

\noindent
Now to extend our discussions further we will consider an excited state by perturbing the pure Lifshitz spacetime. Following the discussion in \cite{Ross:2009ar}, the ground state metric can be perturbed by introducing zero momentum perturbations, which are constant along the boundary direction. These constant perturbations can be broken down into scalar, vector, and tensor components, which is given by\cite{Ross:2009ar}
\begin{equation}
	\tilde{h}_{tt}=-r^{2z}f(r),~~~\tilde{h}_{ti}=-r^{2z}v_{1i}(r)+r^{2}v_{2i}(r),~~~\tilde{h}_{ij}=r^2 k(r)\delta_{ij}+r^2 k_{ij}(r)
\end{equation}   
where 
\begin{equation}
	k_{ij}(r)=\begin{bmatrix}
		t_{d}(r) & t_{0}(r)\\
		t_{0}(r) & -t_{d}(r)
	\end{bmatrix}
\end{equation}
and
\begin{equation}
	a_t =\alpha r^{z}(j(r)+\frac{1}{2}f(r)),~~~a_{i} =\alpha r^{z} v_{1i}(r)~.
\end{equation}
By introducing these perturbations to the pure Lifshitz metric in eq.\eqref{pure lifshitz metric}, one can derive the perturbed Lifshitz metric as follows
\begin{align}
	ds^2&=-r^{2z}(1+f(r))dt^2+r^2 (1+k(r)+t_d (r))dx^2+r^2 (1+k(r)-t_d (r))dy^2+\frac{dr^2}{r^2}\nonumber\\
	&+2[-r^{2z} v_{1x}(r)+r^2 v_{2x}(r)]dt dx+2[-r^{2z} v_{1y}(r)+r^2 v_{2y}(r)]dt dy+2 r^2 h_{xy}(r)dx dy~.\label{perturbed metric}
\end{align}
The above metric can be considered as an excited state of the unperturbed Lifshitz spacetime. Now one can define 
\begin{align}
	&h_{tt}(r)=f(r)\nonumber\\
	&h_{xx}(r)=k(r)+t_d (r)\nonumber\\
	&h_{yy}(r)=k(r)-t_d (r)\nonumber\\
	&a_t=j(r)\label{pertutbations}~.
\end{align}
One should note that the metric in eq.\eqref{perturbed metric}, asymptotically gives Lifshitz spacetime when $h_{tt}(r),h_{xx}(r),\\h_{yy}(r),h_{xy}(r),v_{1x}(r),v_{2x}(r),v_{1y}(r),v_{2y}(r),$ and $a_{t}(r)\to 0$ as $r\to \infty$. The equations of motion determine that the $a_{r}$ component is zero.
In order to determine the linearized action, one needs to substitute eq.\eqref{pertutbations} into eq.\eqref{total action}. By calculating the variation of the action in relation to the boundary metric, one can identify the components of the stress tensor complex. The following equations provide the energy and pressure densities in terms of the functions defined in eq.\eqref{pertutbations}, which will be essential for our subsequent analysis. This reads \cite{Ross:2009ar,Chakraborty:2014lfa}
\begin{align}
	&T_{tt}=-r^{z+2}\left[2r\partial_r k(r)+\alpha^{2}\left(zj(r)+r\partial_r j(r)+\frac{1}{2}r \partial_r f(r)\right)\right]\nonumber\\
	&T_{xx}=-2r^{z+2}\left[(z-1)j(r)-\frac{r}{2}\partial_r f(r)-\frac{r}{2}\partial_r k(r)-\frac{r}{2} (z+2) t_d (r)\right]\nonumber\\
	&T_{yy}=-2r^{z+2}\left[(z-1)j(r)-\frac{r}{2}\partial_r f(r)-\frac{r}{2}\partial_r k(r)+\frac{r}{2} (z+2) t_d (r)\right]~.\label{stess tensor}
\end{align}
By solving the equations of motion in the radial gauge ($h_{r\mu}=0$), 
one can obtain the expression for the functions $j(r),f(r),k(r)$ and $t_{d}(r)$. The solutions for $z=2$ and $z \neq  2$ have quite different forms [\cite{Ross:2009ar,Ross:2011gu,Taylor:2008tg,Andrade:2013wsa}].

\noindent 
For $z=2$, the above functions have the following forms \cite{Ross:2009ar}
\begin{align}
	&j(r)=-\frac{c_1+c_2 \ln r}{r^4}\nonumber\\
	&f(r)=\frac{4c_1-5c_2+4c_2 \ln r}{12 r^4}\nonumber\\
	&k(r)=\frac{4c_1+5c_2+4c_2 \ln r}{24 r^4}\nonumber\\
	&t_d (r)=\frac{t_{d2}}{r^4}\label{fn z=2}~	
\end{align}
and for $z\neq2$
\begin{align}
	&j(r)=-\frac{(z+1)c_1}{(z-1)r^{z+2}}-\frac{(z+1)c_2}{(z-1)r^{\frac{1}{2}(z+2+\beta_z)}}\nonumber\\
	&f(r)=4\frac{1}{(z+2)}\frac{c_1}{r^{z+2}}+2\frac{(5z-2-\beta_z)}{(z+2+\beta_z)}\frac{c_2}{r^{\frac{1}{2}(z+2+\beta_z)}}\nonumber\\
	&k(r)=2\frac{1}{(z+2)}\frac{c_1}{r^{z+2}}-2\frac{(3z-4-\beta_z)}{(z+2+\beta_z)}\frac{c_2}{r^{\frac{1}{2}(z+2+\beta_z)}}\nonumber\\
	&t_d(r)=\frac{t_{d2}}{r^4}\label{fn z not 2}
\end{align}
where $\beta_z ^2=9z^2-20z+20$, $c_1$, $c_2$, $t_{d2}$ are constants of integration. We will now proceed to compute the various information theoretic quantities for the excited state of the Lifshitz spacetime. All the computations are done for the value $z=2$ and as well as for any general value of the dynamical exponent ($z\neq 2$). 

\subsection{Computation of holographic entanglement entropy for excited state}
\noindent
In this subsection, we shall review the computation of the HEE for excited Lifshitz geometry \cite{Chakraborty:2014lfa}.
To compute the expression of the HEE for excited state of Lifshitz spacetime, we will consider a set up similar to that of the ground state of Lifshitz spacetime as mentioned in earlier section \eqref{HEE 2.1}.
To proceed further we will consider a striplike subsystem $A$ on the boundary. The entangling region is taken to be a strip of length $l$ and width $L$ such that $-\frac{l}{2}\leq x\leq\frac{l}{2}$ and  $0\leq y\leq L$. We will vary length along $x$ direction by keeping the width of the subsystem to be fixed. To calculate the area functional of static RT surface, we will first choose a constant time slice, that means we will set $dt=0$. Then we will parameterize the bulk coordinate in terms of the boundary, that is, $r=r(x)$. Taking all these considerations, the expression for the induced metric reads
\begin{equation}
	ds^{2}_{ind}=\left[r^2(1+h_{xx}(r))+\frac{r^{'}(x)^2}{r^2}\right]dx^2+r^2(1+h_{yy}(r))dy^2~.
\end{equation}
Hence, the area functional reads
\begin{align}
	A&=\int_{0}^{L} dy \int_{-l/2}^{l/2} dx \sqrt{g}\nonumber\\
	&=L\int_{-l/2}^{l/2} dx \sqrt{r^4 (1+h_{xx}(r))(1+h_{yy}(r))+r^{'}(x)^2(1+h_{yy}(r))}~.\label{perturbed area functional}
\end{align}
Now taking terms up to $O(h)$ inside the square root in eq.\eqref{perturbed area functional}, we get
\begin{equation}
	A=2L\int_{0}^{l/2} dx \sqrt{r^{'}(x)^2(1+h_{yy}(r))+r^4 (1+h_{xx}(r)+h_{yy}(r))}\label{area functional perturbed}~.
\end{equation}
We can identify the integrand of the above area functional to be the Lagrangian of the form $\mathcal{L}=\mathcal{L}(r,r^{'})$. The above expression of the integrand suggests that $x$ is a cyclic coordinate.  This leads to the following conserved quantity 
\begin{equation}
	H=\frac{-r^4 (1+h_{xx}(r)+h_{yy}(r))}{\sqrt{r^{'}(x)^2(1+h_{yy}(r))+r^4 (1+h_{xx}(r)+h_{yy}(r))}}=constant=c\label{Hamiltonian}~.
\end{equation}
The above constant can be fixed by considering the fact that, at the turning point ($r_{t}$) in the bulk $r^{'}(x)$ vanishes, that is, $r^{'}(x)|_{r_t}=0$. This fixes the above constant to be $c=r^{2}_{t}\sqrt{1+h_{xx}(r_t)+h_{yy}(r_t)}$. This yields the following differential equation
\begin{equation}
	r^{'}(x)=\frac{\sqrt{\frac{r^8}{Q} (1+h_{xx}(r)+h_{yy}(r))^{2}-r^4 (1+h_{xx}(r)+h_{yy}(r))}}{\sqrt{1+h_{yy}(r)}}\label{r prime}~
\end{equation}
where $Q=r^{4}_{t} (1+h_{xx}(r_t)+h_{yy}(r_t))$.\\
Before going to compute the HEE, we now express the subsystem length in terms of bulk coordinate by using the above result. Therefore, the subsystem length in terms of the bulk coordinate reads \cite{Chakraborty:2014lfa}
\begin{align}
	\frac{l}{2}&=\int_{r_t}^{\infty}\frac{dr}{r^2 f(r,r_t)}\left[1-h_{xx}(r)-\frac{1}{2}h_{yy}(r)+\frac{1}{2}(h_{xx}(r_t)+h_{yy}(r_t))\right.\nonumber\\
	&\left.+\frac{(h_{xx}(r_t)+h_{yy}(r_t)-h_{xx}(r)-h_{yy}(r))}{2f^{2}(r,r_t)}\right]\label{length perturbed}
\end{align}
where $f^{2}(r,r_t)=(r/r_t)^2 -1$. 

\noindent
It is important to note that in the above expression $r_t$ is used in place of $r^{(0)}_t$, as we are keeping the subsystem width($l$) to be fixed throughout our calculation the turning point of the RT surface should change due to the metric perturbation. Hence $r_t$ is our new turning point for perturbed Lifshitz spacetime such that $r_t=r^{(0)}_t+\delta r_t$, where $\delta r_t$ is the change in perturbation.

\noindent
Since in our calculation $l$ is kept fixed from eq.\eqref{length pure} we have
\begin{equation}
	\frac{l}{2}=\int_{r_t}^{\infty}\frac{dr}{r^2 f(r,r^{(0)}_t)}\label{length unperturbed}~.
\end{equation}
Substracting the eq(s).(\eqref{length perturbed},\eqref{length unperturbed}) one can easily get the expression for $\delta r_t$ to be\cite{Karar:2017org}

\begin{align}
	\delta r_t &=-N r_t r^{(0)}_t \int_{r_t}^{\infty}\frac{dr}{r^2 f(r,r_t)}\left[h_{xx}(r)+\frac{1}{2}h_{yy}(r)-\frac{1}{2}(h_{xx}(r_t)+h_{yy}(r_t))\right.\nonumber\\
	&\left.-\frac{(h_{xx}(r_t)+h_{yy}(r_t)-h_{xx}(r)-h_{yy}(r))}{2f^{2}(r,r_t)}\right]~.
\end{align}
Now, if we demand $\delta r_t$ to be zero then we get the following relation between the metric perturbations \cite{Allahbakhshi:2013rda}
\begin{equation}
	\resizebox{.92\hsize}{0.025\vsize}{$\frac{(h_{xx}(r_t)+h_{yy}(r_t))}{2f^{2}(r,r_t)}=\frac{1}{[1+f^{2}(r,r_t)]}\left[h_{xx}(r)\left(1+\frac{1}{2f^{2}(r,r_t)}\right)+\frac{h_{yy}(r)}{2}\left(1+\frac{1}{2f^{2}(r,r_t)}\right)\right]$}.\label{perturbation condition}
\end{equation}
The above relation will be useful for deriving the expression for HEE later. In \cite{Allahbakhshi:2013rda}, authors have proposed that insted of working with $r_t =r^{(0)}_t +\delta r_t$, one can use $r_t=r^{(0)}_t$ along with the relation given in eq.\eqref{perturbation condition}. Keeping this in mind, from here on we will use $r^{(0)}_t$ instead of $r_t$ along with the obtained metric perturbation relation in subsequent analysis.
With all these conditions in hand we will proceed to compute a general expression of the RT area functional up to $O(h)$. After substituting the expression of $r^{'}(x)$ from eq.\eqref{r prime} in eq.\eqref{perturbed area functional} and considering terms up to $O(h)$, we get
\begin{equation}
	A=2L\int_{r_t}^{\delta}dr\frac{(1+\frac{1}{2}h_{yy})}{\sqrt{1-\left(\frac{r_t}{r}\right)^{2}}}\left[1+\left(\frac{h_{xx}(r_t)+h_{yy}(r_t)}{2f^{2}(r,r_t)}\right)-\left(\frac{h_{xx}(r)+h_{yy}(r)}{2f^{2}(r,r_t)}\right)\right]
\end{equation}
To proceed further we will use the condition give in eq.\eqref{perturbation condition} in the above equation of the area functional. This yields \cite{Chakraborty:2014lfa}
\begin{align}
	A&=2L\int_{r_{t}^{(0)}}^{\delta} dr \frac{\frac{r^2}{r_{t}^{(0)2}}}{\sqrt{\frac{r^4}{r_{t}^{(0)4}}-1}}+L\int_{r^{(0)}_t}^{\delta}dr \frac{[h_{yy}(r)+(\frac{r^{(0)}_t}{r})^4 h_{xx}(r)]}{\sqrt{1-(\frac{r^{(0)}_t}{r})^4}}\nonumber\\
	&=A^{(0)}+\Delta A~.
\end{align} 
The first term ($A^{(0)}$) in the above equation corresponds to the area of the RT surface for corresponding to the pure Lifshitz spacetime. The expression of $A^{(0)}$ is given by eq.\eqref{A0 25}. On the other hand, the second term refers to the change in the area of RT surface due to perturbation. Hence the change in HEE reads 
\begin{align}
	\Delta S_{HEE} &= \frac{\Delta A}{4 G_4}\nonumber\\
	&=L\int_{r^{(0)}_t}^{\delta}dr \frac{[h_{yy}(r)+(\frac{r^{(0)}_t}{r})^4 h_{xx}(r)]}{\sqrt{1-(\frac{r^{(0)}_t}{r})^4}}~. \label{A in h}
\end{align}

\noindent 
In the above equation, we substitute the expressions of $h_{xx}$ and $h_{yy}$ from eq.\eqref{fn z=2} and eq.\eqref{pertutbations} to obtain the following result of change in HEE for $z=2$
\begin{equation}
	\Delta S^{(2)}_{HEE}= \frac{L\sqrt{\pi} \Gamma(3/4)}{4 G_4 \Gamma(1/4) 24 r^{(0)3}_t}\left[\frac{32 c_1}{5}-\frac{48 t_{d2}}{5}+c_2 \left(\frac{352}{25}-\frac{8\pi}{5}\right)+\frac{32}{5}c_2 \ln r^{(0)}_t\right]~.
\end{equation}
It is to be noted that, to obtain the complete expression of HEE for the excited state of the Lifshitz field theory (or in dual sense, the perturbed Lifshitz geometry), we have to add the HEE associated to the ground state of Lifshitz field theory (or in dual sense, the pure Lifshitz geometry) given in eq.\eqref{A0 25} to the above expression of $\Delta S^{(2)}_{HEE}$. This gives
\begin{align}
	S^{(2)}_{HEE}&=\frac{A}{4 G_4}\nonumber\\
	&=\frac{L}{4 G_4}\left[2\delta -\frac{5}{3}r^{(0)}_t+\frac{\sqrt{\pi} \Gamma(3/4)}{\Gamma(1/4) 24 r^{(0)3}_t}\left\{\frac{32 c_1}{5}-\frac{48 t_{d2}}{5}+c_2 \left(\frac{352}{25}-\frac{8\pi}{5}\right)+\frac{32}{5}c_2 \ln r^{(0)}_t\right\}\right]~.
\end{align}
Now we will obtain the expression of HEE in terms of the subsystem length. To do so, we make use of the relation  given in eq.\eqref{length pure} in the above expression of HEE. This in terns leads to the following
\begin{equation}
	\bar{S}^{(2)}_{HEE}=2\delta -\frac{10}{3Nl}+\frac{N^2 l^3}{192}\left[\tilde{B} + \frac{32}{5}c_2 \ln \frac{2}{Nl}\right]\label{hee z2}~
\end{equation}
where $\bar{S}_{HEE}=\frac{4 G_4}{L} S_{HEE}$, $\tilde{B} = \frac{32 c_1}{5}-\frac{48 t_{d2}}{5}+c_2 (\frac{352}{25}-\frac{8\pi}{5})$.\\
Following a similar procedure, one can also obtain the change in the area of the RT surface for any arbitrary value of the dynamical exponent $z$. This can be done by
substituting the expressions of $h_{xx}$ and $h_{yy}$ (given in eq.(s)(\eqref{pertutbations},\eqref{fn z not 2})) in eq.\eqref{A in h}. This yields 
\begin{equation}
	\Delta A^{(z)}=\frac{L\sqrt{\pi}}{4 r^{(0)(z+1)}_t}\left[\frac{\Gamma(\frac{1+z}{4})}{\Gamma(\frac{3+z}{4})}\left(\frac{2c_1 -t_{d2}}{z+3}\right)+r^{(0)\frac{1}{2} (z+2+\beta_z)}_t \frac{\Gamma(\frac{z+\beta_z}{8})}{\Gamma(\frac{z+4+\beta_z}{8})}\frac{2(4+\beta_z -3z)}{(4+z+\beta_z)}c_2\right]~.
\end{equation}
Hence the change in HEE due to the metric perturbation is given by 
\begin{align}
	\Delta S^{(z)}_{HEE} &= \frac{\Delta A^{(z)}}{4 G_4}\nonumber\\
	&=\frac{L\sqrt{\pi}}{16 G_4 r^{(0)(z+1)}_t}\left[\frac{\Gamma(\frac{1+z}{4})}{\Gamma(\frac{3+z}{4})}\left(\frac{2c_1 -t_{d2}}{z+3}\right)+r^{(0)\frac{1}{2} (z+2-\beta_z)}_t \frac{\Gamma(\frac{z+\beta_z}{8})}{\Gamma(\frac{z+4+\beta_z}{8})}\frac{2(4+\beta_z -3z)}{(4+z+\beta_z)}c_2\right]~.
\end{align}
Furthermore, the full expression of HEE in terms of the subsystem length ($l$) can be written as
\begin{align}
	S^{(z)}_{HEE}=\frac{L}{4 G_4}\big[2\delta -\frac{10}{3Nl}+\frac{\sqrt{\pi}N^{z+1}l^{z+1}}{2^{z+3}}&\biggl\{\frac{\Gamma(\frac{1+z}{4})}{\Gamma(\frac{3+z}{4})}\left(\frac{2c_1 -t_{d2}}{z+3}\right)\nonumber\\
	&+\left(\frac{2}{Nl}\right)^{\frac{1}{2} (z+2-\beta_z)}\frac{\Gamma(\frac{z+\beta_z}{8})}{\Gamma(\frac{z+4+\beta_z}{8})}\frac{2(4+\beta_z -3z)}{(4+z+\beta_z)}c_2\biggr\}\big]~.
\end{align}
In obtaining the above result we have used the result given in eq.\eqref{length pure}, which relates the subsystem length and the turning point. We can recast the above expression of HEE in the following form
\begin{equation}
	\bar{S}^{(z)}_{HEE}=\frac{L}{4 G_4}\left[2\delta -\frac{10}{3Nl}+ A_1 l^{z+1}\biggl\{A_2 + \frac{A_3}{l^{\frac{1}{2} (z+2-\beta_z)}}\biggr\} \right]\label{HEE z not 2}
\end{equation}
where $\bar{S}^{(z)}_{HEE}$, $A_1$, $A_2$ and $A_3$ has the following forms
\begin{equation*}
	\bar{S}^{(z)}_{HEE}=\frac{4G_{4}}{L} S^{(z)}_{HEE}
\end{equation*}
\begin{eqnarray*}
	A_1 = \frac{\sqrt{\pi}N^{z+1}}{2^{z+3}}~~,~~A_2=\frac{\Gamma(\frac{1+z}{4})}{\Gamma(\frac{3+z}{4})}\left(\frac{2c_1 -t_{d2}}{z+3}\right)~,\\
\end{eqnarray*}
\begin{equation*}
	A_3=\left(\frac{2}{N}\right)^{\frac{1}{2} (z+2-\beta_z)}\frac{\Gamma(\frac{z+\beta_z}{8})}{\Gamma(\frac{z+4+\beta_z}{8})}\frac{2(4+\beta_z -3z)}{(4+z+\beta_z)}c_2~.
\end{equation*}
\subsection{Computation of holographic mutual information and EWCS for excited\\ Lifshitz spacetime}
In this subsection we shall calculate HMI and entanglement wedge cross section for the perturbed Lifshitz spacetime. To compute these quantities we will follow the similar procedure to that of the pure Lifshitz spacetime, discussed earlier in subsection \eqref{12}. To proceed further, we will consider two subsystems of equal length $l$, which are separated by the distance $d$. In the limit $d/l<<1$, the expression of HMI can be obtained by using eq.\eqref{mutual ground}. At first we will calculate the HMI for $z=2$ case. To obtain the result of HMI in this case (that is for $z=2$) we will use eq.(s)(\eqref{mutual ground},\eqref{hee z2}). This results in the following expression for HMI
\begin{align}\label{I z2 resacled}
	\bar{I}^{(2)}(A:B)&= \bar{S}_{HEE}(A)+\bar{S}_{HEE}(B)-\bar{S}_{HEE}(A \cup B)\nonumber\\
	&=\frac{10}{3N}\left\{-\frac{2}{l}+\frac{1}{2l+d}+\frac{1}{d}\right\}+\frac{N^2 A_1}{192}\left(2l^3 - (2l+d)^3-d^3\right)\nonumber
	\\&+\frac{c_2 N^2}{30}\left[2l^3 \ln (\frac{2}{Nl})-(2l+d)^3 \ln (\frac{2}{N(2l+d)})-d^3 \ln (\frac{2}{Nd})\right]~
\end{align} 
where $\bar{I}^{(2)}(A:B)=\frac{4G_{4}}{L}I^{(2)}(A:B)$.\\
We denote the change in HMI due to the perturbation is defined as $\Delta \bar{I}(A:B)=\bar{I}(A:B)-\bar{I}^{(0)}(A:B)$, hence for $z=2$, the change in HMI can be written down as 
\begin{align}
	\Delta \bar{I}^{(2)}(A:B)=& \frac{N^2 A_1}{192}\left(2l^3 - (2l+d)^3-d^3\right)\nonumber
	\\&+\frac{c_2 N^2}{30}\left[2l^3 \ln (\frac{2}{Nl})-(2l+d)^3 \ln (\frac{2}{N(2l+d)})-d^3 \ln (\frac{2}{Nd})\right]~.
\end{align}
Now we will proceed to compute the HMI for arbitrary value of the dynamical exponent. To do so we will consider the similar set up to that of $z=2$. Therefore, the expression of HMI for two subsystems of equal length $l$ and which are separated by distance $d$ can be obtained by using
eq.(s)(\eqref{mutual ground},\eqref{HEE z not 2}). Hence, in this setup the expression of HMI for arbitrary value of dynamical exponent reads 
\begin{align}\label{Iz}
	I^{(z)}(A:B)=&\frac{L}{4G_4}\left[\frac{10}{3N}\biggl\{\frac{1}{2l+d}+\frac{1}{d}-\frac{2}{l}\biggr\}+\biggl\{2l^{z+1}-(2l+d)^{z+1}-d^{z+1}\biggr\}A_1 A_2 \right. \nonumber \\&\left. +\biggl\{2l^{\left(\frac{z+\beta_z}{2}\right)}-(2l+d)^{\left(\frac{z+\beta_z}{2}\right)}-d^{\left(\frac{z+\beta_z}{2}\right)}\biggr\}A_1 A_3 \right]~.
\end{align}
Now we will proceed to compute the entanglement wedge cross section for the perturbed Lifshitz geometry in (3+1)-dimensions. By following the approach we have shown in subsection \eqref{12}, we consider two subsystems of equal length $l$ and which are separated by a distance $d$, where $d/l<<1$. As shown earlier, we have to calculate the vertical constant $x$ hypersurface with minimal area which splits the entanglement wedge $M_{AB}$ into two domains corresponding to $A$ and $B$. The induced metric for a constant vertical $x$ hypersurface on a constant time slice can be found from eq.\eqref{perturbed metric}, and this reads
\begin{equation}
	ds^2_{ind} = r^2 (1+h_{yy}(r)) + \frac{dr^2}{r^2}~. 
\end{equation}
The above induced metric leads to the following expression of EWCS 
\begin{align}
	E_W &= \frac{1}{4 G_4}\int_{0}^{L}dy \int_{r^{(0)}_{t}(2l+d)}^{r^{(0)}_{t}(d)} dr \sqrt{1+h_{yy}(r)}\nonumber\\
	&=\frac{L}{4 G_4} \int_{r^{(0)}_{t}(2l+d)}^{r^{(0)}_{t}(d)} dr (1+\frac{1}{2} h_{yy}(r))~.
\end{align}
We will now define a quantity $\bar{E}_{W}=\frac{4 G_4}{L}E_W$. Hence 
\begin{equation}
	\bar{E}_W = \int_{r^{(0)}_{t}(2l+d)}^{r^{(0)}_{t}(d)} dr (1+\frac{1}{2} h_{yy}(r))\label{EWCS integral form}~.
\end{equation}
To calculate the value of EWCS for $z=2$ scenario, we substitute the form of $h_{yy}$ by using eq.\eqref{pertutbations} and eq.\eqref{fn z=2}. Therefore, the expression for EWCS for $z=2$ reads
\begin{align}
	\bar{E}_{W}^{(2)} &= [r^{(0)}_{t}(d)-r^{(0)}_{t}(2l+d)]\nonumber\\& + \frac{1}{6} \left(\frac{c_1}{6}+\frac{11}{72}c_2- t_{d2}\right)\left(\frac{1}{r^{(0)3}_{t}(2l+d)}-\frac{1}{r^{(0)3}_{t}(d)}\right)
	+\frac{c_2}{36}\left(\frac{\ln r^{(0)}_{t}(2l+d)}{r^{(0)3}_{t}(2l+d)}-\frac{\ln r^{(0)}_{t}(d)}{r^{(0)3}_{t}(d)}\right)~.
\end{align}
The next step is to express the above expression of EWCS in terms of the subsystem length $l$ and the distance of separation $d$. In order to do this, we use eq.\eqref{length pure} in the above result. This in turn gives us the following expression of EWCS in terms of the subsystem length $(l)$ and separation distance $(d)$
\begin{align}\label{EW z2 rescaled}
	\bar{E}_{W}^{(2)}=\frac{2}{N}\left[\frac{1}{d}-\frac{1}{(2l+d)}\right]&+\frac{N^3}{48}\left(\frac{c_1}{6}+\frac{11}{72}c_2- t_{d2}\right)\left[(2l+d)^3 -d^3\right]\nonumber\\
	&+\frac{c_2 N^3}{288} \left[(2l+d)^3 \ln \left(\frac{2}{N(2l+d)}\right)- d^3 \ln \left(\frac{2}{Nd}\right) \right]~.
\end{align}
With all these expressions in hand we will now compute the critical separation length $d_c$ at which HMI vanishes for fixed choice of subsystem length. It was argued in \cite{Takayanagi:2017knl} that, at a critical separation HMI becomes zero, which implies that entanglement wedge becomes disconnected. Some recent works in this direction can be found in \cite{Basu:2021awn,Sahraei:2021wqn,Chowdhury:2021idy,RoyChowdhury:2023iyr,Liu:2021rks}. This discontinuity represent a phase transition of EWCS from the connected phase to the disconnected phase. To calculate the critical separation length ($d_{c}$), we have to solve the following equation for $d_{c}$
\begin{align}
	&\frac{10}{3N}\left\{-\frac{2}{l}+\frac{1}{2l+d_{c}}+\frac{1}{d_{c}}\right\}+\frac{N^2 A_1}{192}\left(2l^3 - (2l+d_{c})^3-d^{3}_{c}\right)\nonumber
	\\&+\frac{c_2 N^2}{30}\left[2l^3 \ln (\frac{2}{Nl})-(2l+d_{c})^3 \ln (\frac{2}{N(2l+d_{c})})-d^{3}_{c} \ln (\frac{2}{Nd_{c}})\right]=0~
\end{align}
where $d_{c}$ is the critical separation at which HMI vanishes. The above equation suggests that the critical separation $d_c$ depends on subsystem length and the constants. This means as long as the separation distance between two subsystems is smaller than the critical distance, that is, $d<d_c$, the entanglement wedge associated to the full system is in the connected phase and for separation distance greater than the critical separation, the entanglement wedge is in the disconnected phase.

\begin{figure}[!h]
	\centering
	\begin{subfigure}[t]{0.5\textwidth}
		\centering
		\includegraphics[width=8.5cm]{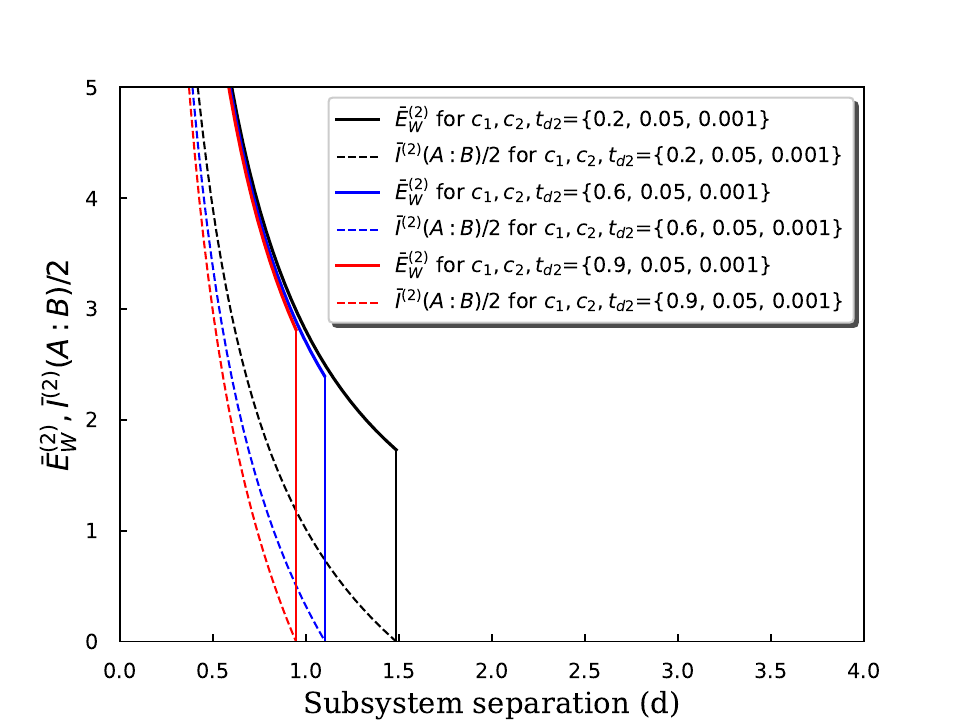}
		\caption{Plot for different values of the constants (for $z=2$)}
		\label{HMI EWCS z2 different const}
	\end{subfigure}%
	~
	\begin{subfigure}[t]{0.5\textwidth}
		\centering
		\includegraphics[width=8.5cm]{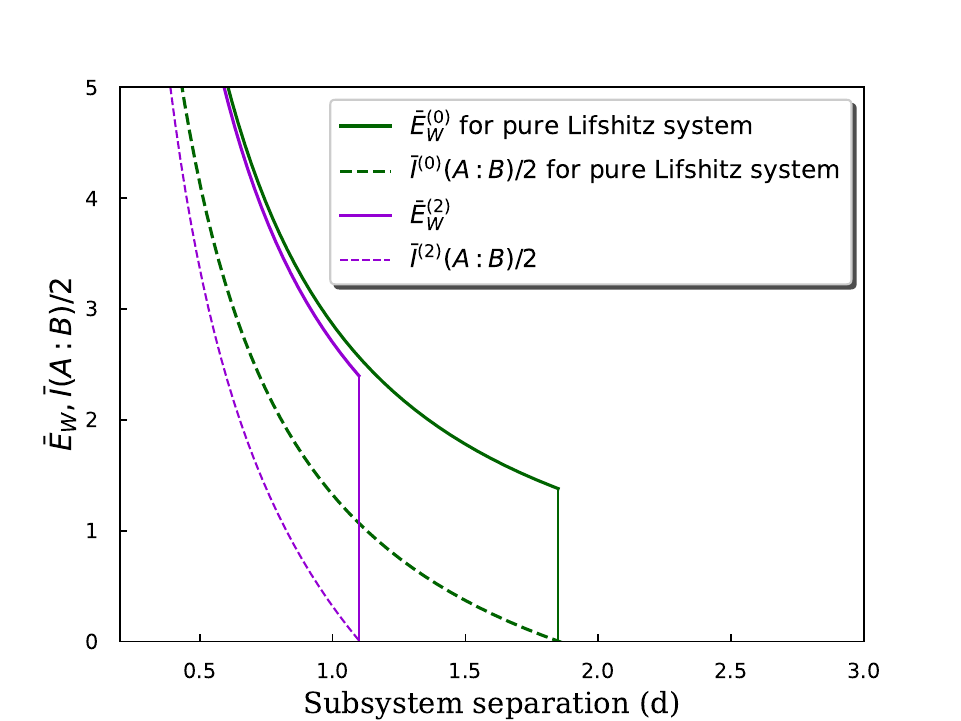}
		\caption{Comparison between the results of ground state and excited state}
		\label{HMI EWCS z2 pure compare}
	\end{subfigure}
	\caption{Variation of EWCS and HMI for two disjoint subsystems (with length $l=3$) with respect to their separation distance. In the left panel of the above Figure, we have represented the variation of EWCS and HMI with the separation distance between two subsystems. To do this plot we have chosen three sets of constants by following the inequality given in eq.\eqref{inequality const}. In the right panel of the Figure, we have compared the results of EWCS and HMI for excited state (for $z=2$) and ground state Lifshitz spacetime.}
\end{figure}

\noindent
Now we will graphically represent the results of EWCS and HMI. In the present case, it is a bit tricky to plot these results. The underlying reason can be explained by the following argument. To represent these results graphically we have to know the value of the constants ($c_1$, $c_2$ and $t_{d2}$) appearing in the expression of EWCS and HMI. To get rid of this issue we have found an allowed range of the constants for which the following inequality holds $E_{W}(A:B)\geq \frac{I(A:B)}{2}$ \cite{Takayanagi:2017knl,Terhal_2002,Nguyen:2017yqw}. We observe that the mentioned inequality is satisfied as long as the following condition is valid
\begin{equation}
	\frac{1}{3d}-\frac{10}{3l}-\frac{11}{3(2l+d)}+\alpha (d,l) c_1 +\beta (d,l) c_2 +
	\gamma (d,l) t_{d2}\geq 0 \label{inequality const}
\end{equation}
where
\begin{equation*}
	\alpha (d,l)= \frac{N^{3}}{720}\left(24 d^3 +4l^3 (18+5N)+3dl(24+5N)(d+2l)\right)
\end{equation*}
\begin{align*}
	\beta (d,l)=&\frac{N^3}{43200}\left(4l^3 (475N+54(44-5\pi)+3dl(475N+72(44-5\pi))(d+2l)+72d^3 (44-5\pi))\right.\\ &\left. -1440~l^3 ln\left(\frac{2}{Nl}\right)+30\left(d^3 (24-5N)ln\left(\frac{2}{Nd}\right)+(d+2l)^3 (24+5N)ln\left(\frac{2}{N(d+2l)}\right)\right) \right)
\end{align*}
\begin{equation*}
	\gamma (d,l)=-\frac{N^3}{120}\left(6d^3 +2l^3 (9+10N)+3dl(6+5N)(d+2l)\right)~.
\end{equation*}
\vskip 0.4cm
\noindent
It is to be noted that the above condition must hold for every value of the separation distance $d$, corresponding to a give subsystem length $l$.
\vskip 0.2cm
\noindent
In the Fig.\eqref{HMI EWCS z2 different const}, we have choosen three different sets of constants (obeying the inequality given in eq.\eqref{inequality const}) and for each set we have plotted the variation HMI and EWCS with respect to the subsystem separation ($d$) by following eq.(s)(\eqref{EW z2 rescaled},\eqref{I z2 resacled} respectively (by keeping the subsystem length to be fixed). The plot in black color is for $c_1 $, $c_2$ and $t_{d2}$=${0.2,0.05,0.001}$, the blue one is for $c_1 $, $c_2$ and $t_{d2}$=${0.6,0.05,0.001}$ and the red one is for $c_1 $, $c_2$ and $t_{d2}$=${0.9,0.05,0.001}$, respectively. Fig.\eqref{HMI EWCS z2 pure compare} shows a graphical comparison between the results of HMI and EWCS for excited state and ground state of Lifshitz spacetime. The plot in green is for pure Lifshitz spacetime and the plot in purple represents the results for excited state corresponding to $z=2$ with the choice of the constants $c_1 $, $c_2$ and $t_{d2}$=${0.6,0.05,0.001}$. This plot suggests that the mutual information for ground state Lifshitz spacetime vanishes later compared to the excited state associated with $z=2$. Hence, the entanglement wedge becomes disconnected earlier for excited state compared to ground state (for $z=2$) of the Lifshitz spacetime.
\vskip 0.2cm
\noindent
Next, we will determine the change in the EWCS resulting from the perturbation in the pure Lifshitz spacetime. This we define as
\begin{alignat}{2}
	\Delta \bar{E}_{W}^{(2)}&=\bar{E}_{W}^{(2)}-\bar{E}^{(0)}_{W}\nonumber\\
	=&\frac{N^3}{48}\left(\frac{c_1}{6}+\frac{11}{72}c_2- t_{d2}\right)\left[(2l+d)^3 -d^3\right]+\frac{c_2 N^3}{288} \left[(2l+d)^3 \ln \left(\frac{2}{N(2l+d)}\right)- d^3 \ln \left(\frac{2}{Nd}\right) \right]\label{change EW z2}~.
\end{alignat}
\noindent
We shall next proceed to compute to compute the EWCS for arbitrary value of the dynamical exponent, that is for $z \neq 2$. In this scenario we can compute EWCS by substituting the form of $h_{yy}$ in eq.\eqref{EWCS integral form} from eqs.(\eqref{pertutbations},\eqref{fn z not 2}). This results in the following expression of EWCS for arbitrary value of the dynamical exponent in terms of the turning point 
\begin{align}
	\bar{E}_{W}^{(z)}=\left[r_{t}(d)-r_{t}(2l+d)\right]&+ \left(\frac{2c_1}{z+2}-t_{d2}\right)\frac{1}{2(z+1)}\left[\frac{1}{r^{z+1}_{t}(2l+d)}-\frac{1}{r^{z+1}_{t}(d)}\right]\nonumber\\
	&-\frac{(3z-4-\beta_z)}{(z+2+\beta_z)}\frac{2c_2}{(z+\beta_z)}\left[\frac{1}{r^{\frac{z+\beta_z}{2}}_{t}(2l+d)}-\frac{1}{r^{\frac{z+\beta_z}{2}}_{t}(d)}\right]~.
\end{align}
Now to rewrite the expression of EWCS in terms of subsystem length, we will use the relation between the subsystem length and the turning point given in eq.\eqref{length pure} in the above result. This yields
\begin{align}\label{EWCS z}
	\bar{E}_{W}^{(z)}=\frac{2}{N}\left[\frac{1}{d}-\frac{1}{2l+d}\right]&+ \left(\frac{c_1}{z+2}-\frac{t_{d2}}{2}\right)\frac{N^{z+1}}{2^{z+1}(z+1)}\left[(2l+d)^{z+1}-d^{z+1}\right]\nonumber\\
	&-2c_2\frac{(3z-4-\beta_z)}{(z+2+\beta_z)}\frac{N^{\frac{z+\beta_z}{2}}}{2^{\frac{z+\beta_z}{2}}(z+\beta_z)}\left[(2l+d)^{\frac{z+\beta_z}{2}}-d^{\frac{z+\beta_z}{2}}\right]~.
\end{align}
Similar to the $z=2$ scenario, we will now compute the critical separation at which HMI vanishes for a fixed choice of subsystem length. To calculate the critical separation for an arbitrary dynamical exponent, we have to solve the following equation for $\tilde{d}_{c}$ 
\begin{align}
	&\left[\frac{10}{3N}\biggl\{\frac{1}{2l+\tilde{d}_{c}}+\frac{1}{\tilde{d}_{c}}-\frac{2}{l}\biggr\}+\biggl\{2l^{z+1}-(2l+\tilde{d}_{c})^{z+1}-\tilde{d}_{c}^{z+1}\biggr\}A_1 A_2 \right. \nonumber \\&\left. +\biggl\{2l^{\left(\frac{z+\beta_z}{2}\right)}-(2l+\tilde{d}_{c})^{\left(\frac{z+\beta_z}{2}\right)}-\tilde{d}_{c}^{\left(\frac{z+\beta_z}{2}\right)}\biggr\}A_1 A_3 \right]=0~
\end{align}
where $\tilde{d}_{c}$ is the critical separation at which HMI vanishes. The above equation suggests that the critical separation depends on  constants appearing in the above equation and dynamical scaling exponent ($z$), for a given choice of subsystem length. This means that as long as the subsystem separation is smaller than the critical distance, that is, $d<\tilde{d}_{c}$ the entanglement wedge is in the connected phase and when $d>\tilde{d}_{c}$ the entanglement wedge is in the disconnected phase. \\
Now we will graphically represent the variation of HMI and EWCS for an arbitrary $z$. As discussed earlier, it is a bit tricky to plot the results graphically because of the presence of the constants ($c_1$, $c_2$ and $t_{d2}$) in results of EWCS and HMI. To get rid of this problem we can again derive an allowed range of constants using the inequality $E_{W}(A:B)\geq \frac{I(A:B)}{2}$. 
\begin{figure}[!h]
	\centering
	\begin{subfigure}[t]{0.5\textwidth}
		\centering
		\includegraphics[width=8.5cm]{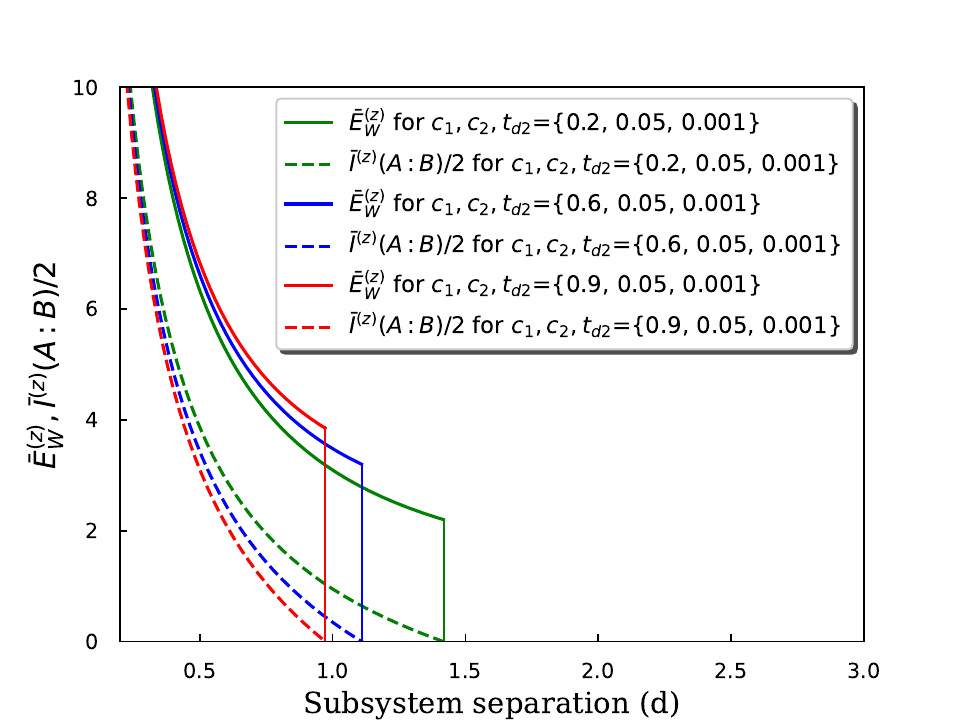}
		\caption{Plot for different values of the constants \\(for $z=3$)}
		\label{same z different constants}
	\end{subfigure}%
	~ 
	\begin{subfigure}[t]{0.5\textwidth}
		\centering
		\includegraphics[width=8.5cm]{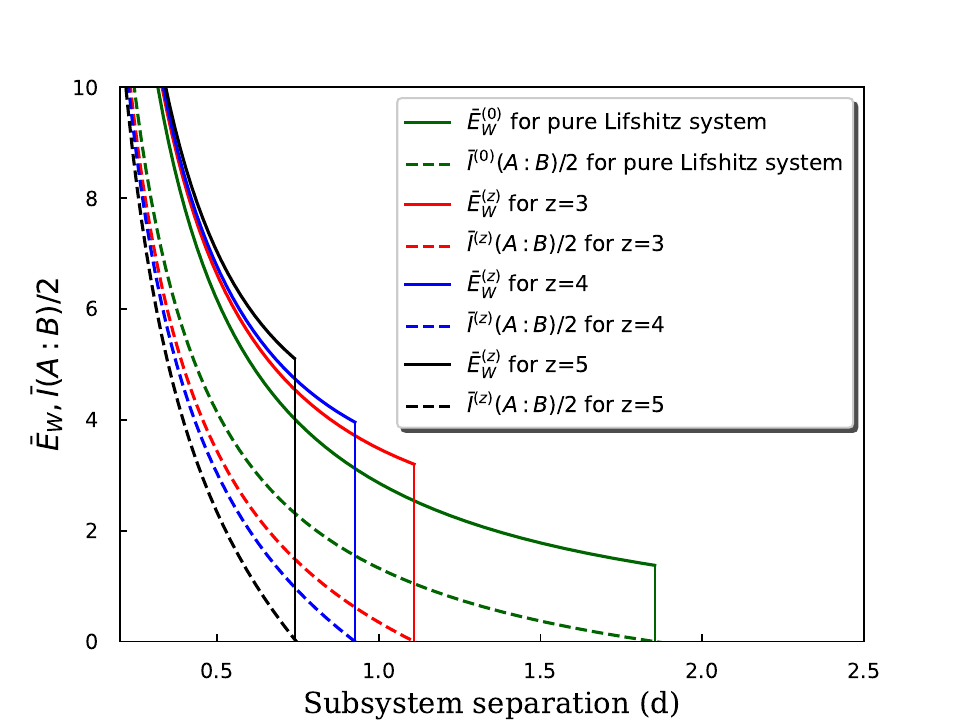}
		\caption{comparison between the results of ground and \\excited state}
		\label{comaprison pure vs z}
	\end{subfigure}\\[1ex]
	\caption{In the above Figures, we have shown (a) variation of EWCS and HMI for different values of constants, (b)  comparison between the variation of EWCS and HMI for pure and excited states for $z=3,4$ and $5$.}
\end{figure}
\noindent
Hence the inequality satisfied by $c_1$, $c_2$ and $t_{d2}$ for arbitrary value of dynamical scaling exponent reads 
\begin{align}
	\frac{1}{3d}-\frac{10}{3l}-\frac{11}{3(2l+d)}+\left(\frac{c_1}{z+2}-\frac{t_{d2}}{2}\right)\frac{N^{z+1}}{2^{z+1}(z+1)}((2l+d)^{z+1}-d^{z+1})\nonumber\\ -\frac{2c_{2}(3z-4-\beta_{z})}{(z+2+\beta_{z})(z+\beta_{z})}\frac{N^{\frac{z+\beta_{z}}{2}}}{2^{\frac{z+\beta_{z}}{2}}}\left((2l+d)^{\frac{z+\beta_{z}}{2}}-d^{\frac{z+\beta_{z}}{2}}\right)-\frac{A_{1}A_{2}}{2}\left((2l+d)^{z+1}-d^{z+1}\right)\nonumber\\ -\frac{A_{2}A_{3}}{2}\left((2l+d)^{\frac{z+\beta_{z}}{2}}-d^{\frac{z+\beta_{z}}{2}}\right)\geq 0 \label{inequality z}
\end{align}
In Fig.\eqref{same z different constants}, by following the eq.(s)(\eqref{EWCS z},\eqref{Iz}), we have plotted the variation of HMI and EWCS with respect to the subsystem separation $d$, to do so we have chosen three different sets of constants by following eq.\eqref{inequality z}. The plot in green color is for $c_1 $, $c_2$ and $t_{d2}$=${0.2,0.05,0.001}$, the blue one is for $c_1 $, $c_2$ and $t_{d2}$=${0.6,0.05,0.001}$ and the red one is for $c_1 $, $c_2$ and $t_{d2}$=${0.9,0.05,0.001}$, respectively. Fig.\eqref{comaprison pure vs z} shows a graphical comparison between the results of HMI and EWCS for excited state (corresponding to $z=3,4$ and $5$) and ground state of Lifshitz spacetime. The plot in green is for pure Lifshitz spacetime and the plots in red, blue and black are for $z$ values 3,4 and 5 respectively. To do this plot we have chosen $c_1 $, $c_2$ and $t_{d2}$=${0.6,0.05,0.001}$. This implies for any arbitrary value of dynamical exponent the mutual information for excited state vanishes earlier compared to the ground state of Lifshitz spacetime. From the plot it can also be seen that for higher values of the dynamical exponent the critical separation distance decreases. Hence the phase transition between connected and disconnected phases of EWCS occurs earlier for higher values of dynamical exponent.  
This in terns means that the entanglement wedge associated with the excited state (for arbitrary value of dynamical exponent) becomes disconnected for smaller value of the critical separation compared to the pure Lifshitz spacetime.\\

\noindent
Hence, the change in EWCS in this case is given by 
\begin{align}\label{change in EW for z}
	\Delta\bar{E}_{W}^{(z)}=\bar{E}_{W}^{(z)}-\bar{E}_{W}^{(0)}=&\left(\frac{c_1}{z+2}-\frac{t_{d2}}{2}\right)\frac{N^{z+1}}{2^{z+1}(z+1)}\left[(2l+d)^{z+1}-d^{z+1}\right]\nonumber\\
	&-2c_2\frac{(3z-4-\beta_z)}{(z+2+\beta_z)}\frac{N^{\frac{z+\beta_z}{2}}}{2^{\frac{z+\beta_z}{2}}(z+\beta_z)}\left[(2l+d)^{\frac{z+\beta_z}{2}}-d^{\frac{z+\beta_z}{2}}\right]~.
\end{align}

\subsection{Entanglement Negativity}
\begin{figure}[!h]
	\centering
	\begin{subfigure}[t]{0.5\textwidth}
		\centering
		\includegraphics[width=8.5cm]{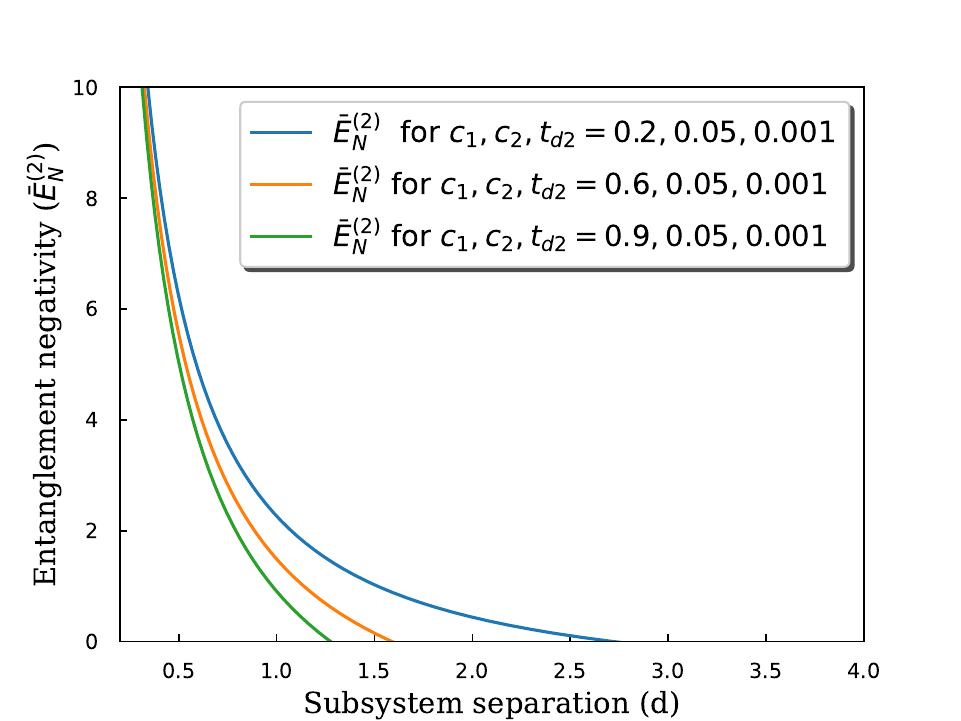}
		\caption{Different values of constants}
		\label{EN z2 different const}
	\end{subfigure}%
	~ 
	\begin{subfigure}[t]{0.5\textwidth}
		\centering
		\includegraphics[width=8.5cm]{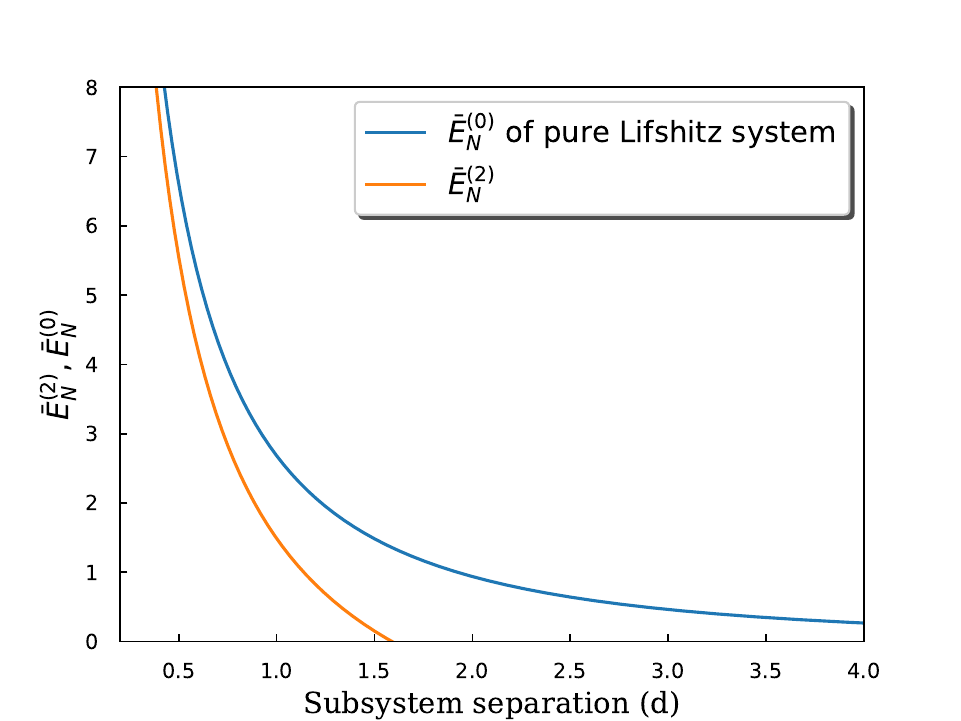}
		\caption{comparison with the pure system}
		\label{EN z2 pure compare}
	\end{subfigure}
	\caption{Variation of entanglement negativity for two disjoint subsystems (with length $l$) with respect to their separation distance ($z=2$ scenario). }
\end{figure}

\begin{figure}[!h]
	\centering
	\begin{subfigure}[t]{0.5\textwidth}
		\centering
		\includegraphics[width=8.5cm]{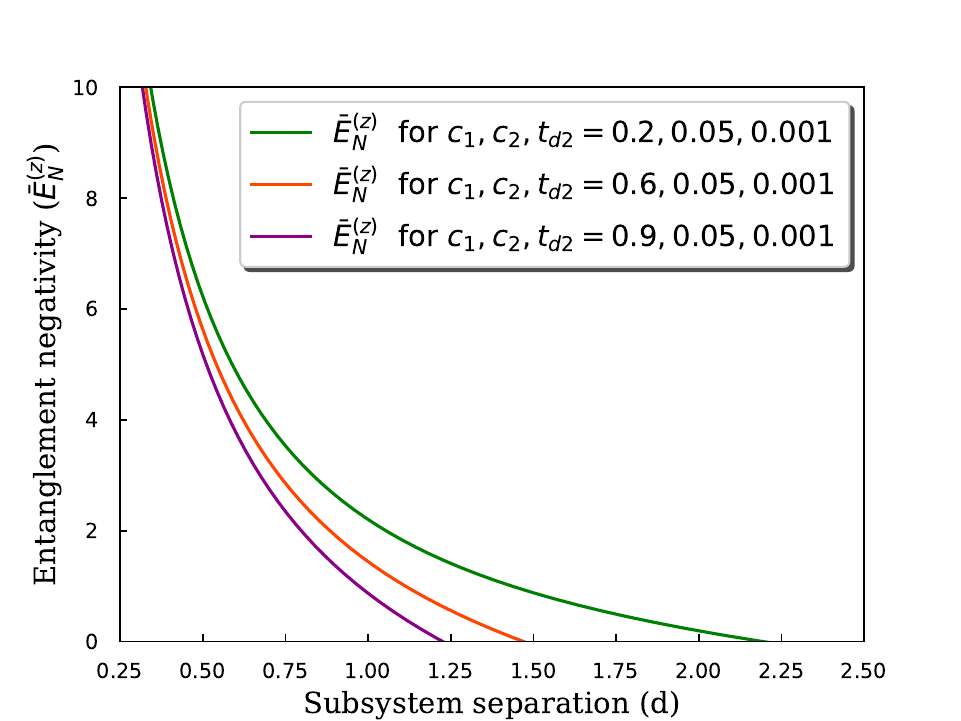}
		\caption{Plots for different values of the constants for fixed dynamical exponent ($z=3$)}
		\label{En different const}
	\end{subfigure}%
	~ 
	\begin{subfigure}[t]{0.5\textwidth}
		\centering
		\includegraphics[width=8.5cm]{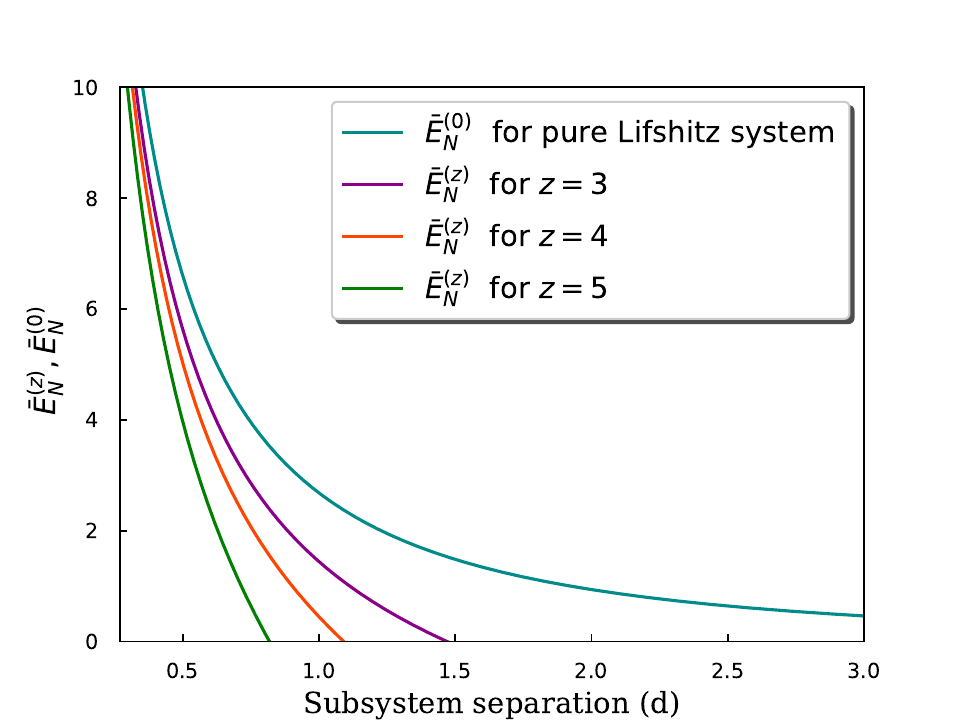}
		\caption{comparison between pure and excited system}
		\label{En comp vs z}
	\end{subfigure}\\[1ex]
	\caption{In the above figure we have shown the variation of entanglement negativity for two disjoint subsystems with the distance of separation. In particular, in Fig.(6a) we have shown the variation of entanglement negativity with subsystem separation for three different sets of constants with fixed dynamical scaling exponent $z=3$ and subsystem length $l=3$. The plot in green is for $c_1 $, $c_2$ and $t_{d2}$=${0.2,0.05,0.001}$, the plot in orange is for $c_1 $, $c_2$ and $t_{d2}$=${0.6,0.05,0.001}$ and the plot in purple is for $c_1 $, $c_2$ and $t_{d2}$=${0.9,0.05,0.001}$. In Fig.(6b) we have shown a graphical comparison between the results of entanglement negativity for pure and excited state (for $z=3,4$ and $5$) of Lifshitz spacetime. The plot in green color represents the result of pure Lifshitz spacetime and the plots in purple, orange and green are for the excited scenario with $z=3,4$ and $5$ respectively. The entanglement negativity for excited state is plotted for $c_1 $, $c_2$ and $t_{d2}$=${0.6,0.05,0.001}$. }
	\label{Enz graph}
\end{figure}

In this subsection we will compute the entanglement negativity for excited state of Lifshitz spacetime holographically. We will compute $E_N$ for two different setups, namely, for adjacent and disjoint subsystems. First we will consider two adjacent subsystems of length $l_1$ and $l_2$ at the boundary. In this setup the entanglement negativity for excited state of Lifshitz spacetime corresponding to $z=2$ can be obtained by using eq.(s)(\eqref{hee z2},\eqref{negetivity attached}). This results the following expression of entanglement negativity for $z=2$
\begin{align}
	\bar{E}^{(2)}_{N_{adj}}=&\frac{3}{4} \left[2\delta- \frac{10}{3N}\left(\frac{1}{l_1}+\frac{1}{l_2}-\frac{1}{l_1+l_2}\right)+\frac{N^2 B}{192}\left(l^{3}_{1}+l^{3}_{2}-(l_1+l_2)^3\right)\right.\nonumber\\
	&\left. -\frac{N^2 c_2}{30}\left(l^{3}_{1}\ln l_{1}+l^{3}_{2}\ln l_{2}-(l_1+l_2)^3 \ln(l_1 + l_2)\right) \right]~.
\end{align} 
where $B=\tilde{B}+\frac{32}{5}c_{2}\ln\left(\frac{2}{N}\right)$. The above expression suggests that the entanglement negativity for adjacent subsystems contains a divergent piece. Now if we consider the length of two subsystems are equal, that is, $l_1 = l_2 =l$, the expression of the entanglement negativity becomes
\begin{equation}
	\bar{E}^{(2)}_{N_{adj}}=\frac{3}{4}\left[2\delta -\frac{5}{Nl}-\frac{N^2 B}{32}l^3+\frac{N^2 c_2}{15}l^{3}\ln 16l^3\right]\label{E_N adjacent z2}~.
\end{equation}
We will now calculate the change in entanglement negativity due to the perturbation, this is computed as
\begin{equation}
	\Delta \bar{E}^{(2)}_{N_{adj}}=\bar{E}^{(2)}_{N_{adj}}-\bar{E}^{(0)}_{N_{adj}}=-\frac{3}{4}\left[\frac{N^2 B}{32}l^3-\frac{N^2 c_2}{15}l^{3}\ln 16l^3\right]\label{change EN adjacent z2}~.
\end{equation}
Similar to the $z=2$ scenario now we will proceed to compute the entanglement negativity for arbitrary value of dynamical scaling exponent. To do so we will substitute the value of HEE from eq.\eqref{HEE z not 2} in the formula of entanglement in eq.\eqref{negetivity attached}. This yields
\begin{align}
	\bar{E}^{(z)}_{N_{adj}}=\frac{3}{4}\left[2\delta +\frac{10}{3N}\left(\frac{1}{l_1 + l_2}-\frac{1}{l_1}-\frac{1}{l_2}\right)\right.&\left.+A_1 A_2 \left(l^{z+1}_{1}+l^{z+1}_{1}-(l_1 +l_2)^{z+1}\right)\right.\nonumber\\&\left. +A_1 A_3 \left(l^{\frac{z+\beta_z}{2}}_{2}+l^{\frac{z+\beta_z}{2}}_{1}-(l_1 + l_2)^{\frac{z+\beta_z}{2}}_{1}\right)\right]~
\end{align} 
for two subsystems of equal width the above expression becomes 
\begin{align}
	\bar{E}^{(z)}_{N_{adj}}=\frac{3}{4}\left[2\delta -\frac{5}{Nl}+2A_1 A_2 l^{z+1}(1-2^{z})+2A_1 A_3 l^{\frac{z+\beta_z}{2}}(1-2^{\frac{z+\beta_z -2}{2}})\right]~.\label{E_N adjacent z not 2}
\end{align}
Similar to the $z=2$ case, the entanglement negativity for two adjacent subsystems, with arbitrary value of dynamical scaling exponent also contains a divergent piece. 
The change in entanglement negativity (for adjacent scenario) for arbitrary value of dynamical exponent is given by  
\begin{equation}
	\Delta \bar{E}^{(z)}_{N_{adj}}=\frac{3}{4}\left[2A_1 A_2 l^{z+1}(1-2^{z})+2A_1 A_3 l^{\frac{z+\beta_z}{2}}(1-2^{\frac{z+\beta_z -2}{2}})\right]~.\label{change EN adjoint znot2}
\end{equation}
Next we will proceed to compute the entanglement negativity for disjoint subsystems. 
The entanglement negativity of two disjoint subsystems when $z=2$ can be calculated following the eq.(s)(\eqref{negetivity disjoint},\eqref{hee z2}), and this reads
\begin{align}
	\bar{E}^{(2)}_{N_{dis}}&=\frac{10}{3N}\biggl\{\frac{1}{l_1 +l_2 +d}+ \frac{1}{d}-\frac{1}{l_1 +d}-\frac{1}{l_2 + d}\biggr\}\nonumber
	+\frac{N^2 B}{192}\biggl\{(l_1 +d)^3 + (l_2 +d)^3-(l_1 +l_2 +d)^3 -d^3\biggr\} \\&+\frac{N^2 c_2}{30} \biggl\{(l_1 + l_2 +d)^3 \ln(l_1 + l_2 +d)
	+d^3 \ln d-(l_1 + d)^3 \ln(l_1 +d)\nonumber\\&-(l_2 +d)^3 \ln(l_2 +d)\biggr\}~.
\end{align}
The above equation suggests that entanglement negativity for disjoint subsystems does not contain any divergent piece. If we consider $l_1 =l_2 =l$, the above expression becomes
\begin{align}
	\bar{E}^{(2)}_{N_{dis}}&=\frac{10}{3N}\biggl\{\frac{1}{2l +d}+ \frac{1}{d}-\frac{2}{l +d}\biggr\}\nonumber
	+\frac{N^2 B}{192}\biggl\{2(l +d)^3 -( 2l+d)^3 -d^3\biggr\} \\&+\frac{N^2 c_2}{30} \biggl\{(2l+d)^3 \ln(2l +d)
	+ d^3 \ln d-2(l + d)^3 \ln(l +d)\biggr\}\label{E_N disjoint z2}~.
\end{align}
Now to understand the effect on metric perturbation on entanglement negativity (for dynamical exponent $z=2$), we will proceed to represent the variation of entanglement negativity with respect to the subsystem separation distance $d$ graphically. The constants in the above expression of entanglement negativity are chosen by following the inequality in eq.\eqref{inequality const}, to plot the results we will follow the same procedure as discussed earlier for EWCS. In Fig.\eqref{EN z2 different const}, we have chosen three different sets of constants and for each set we have shown the variation of entanglement negativity with respect to the subsystem separation ($d$) by following eq.\eqref{E_N disjoint z2}. The plot in blue is for $c_1 $, $c_2$ and $t_{d2}$=${0.2,0.05,0.001}$, the orange one is for $c_1 $, $c_2$ and $t_{d2}$=${0.6,0.05,0.001}$ and the green one is for $c_1 $, $c_2$ and $t_{d2}$=${0.9,0.05,0.001}$. 
The plots suggests that entanglement negativity for two disjoint subsystems vanishes for a particular value of the separation distance, $d^{'}_{c}$. Hence the entanglement negativity measures correlation between two disjoint subsystems even when they are not in the connected phase, this is due to the fact that entanglement negativity vanishes at large values of subsystem separation compared to HMI and EWCS.
In Fig.\eqref{EN z2 pure compare} we have shown a graphical comparison between the result of entanglement negativity for excited and ground state of Lifshitz spacetime. The plot in orange represents the result for excited state corresponding to $z=2$ with the choice of the constants $c_1 $, $c_2$ and $t_{d2}$=${0.6,0.05,0.001}$ and the plot in blue represents the pure Lifshitz spacetime. This the plot shows that, entanglement negativity for the excited state corresponding to $z=2$ vanishes earlier compared to that of the pure Lifshitz spacetime. 
\noindent
Similarly to the adjoint scenario, we will calculate the variation in $E_N$ resulting from the perturbation for two separate subsystems. The change in entanglement negativity reads
\begin{align}
	\Delta \bar{E}^{(2)}_{N_{dis}}&=\frac{3}{4}\left[\frac{N^2 B}{192}\biggl\{2(l +d)^3 -( 2l+d)^3 -d^3\biggr\}\right. \nonumber\\&\left.+\frac{N^2 c_2}{30} \biggl\{(2l+d)^3 \ln(2l +d)
	+ d^3 \ln d-2(l + d)^3 \ln(l +d)\biggr\}\right]\label{change EN disjoint z2}~.
\end{align}
To compute the entanglement negativity of two disjoint subsystems for arbitrary value of the dynamical exponent (except $z=2$) we will use the eq.(s)(\eqref{negetivity disjoint},\eqref{HEE z not 2}). This yields
\begin{align}
	\bar{E}^{(z)}_{N_{dis}}=&\frac{3}{4}\left[\frac{10}{3N}\left(\frac{1}{l_1 +l_2 +d}+\frac{1}{d}-\frac{1}{l_1  +d}-\frac{1}{l_2 +d}\right)\right.\nonumber\\ &\left. +A_1 A_2 \left((l_1 +d)^{z+1}+(l_2 +d)^{z+1}-(l_1 +l_2 +d)^{z+1}-d^{z+1}\right)\right.\nonumber\\&\left. +A_1 A_3 \left((l_1 +d)^{\frac{z+\beta_z}{2}}+(l_2 +d)^{\frac{z+\beta_z}{2}}-(l_1 +l_2 +d)^{\frac{z+\beta_z}{2}}-d^{\frac{z+\beta_z}{2}}\right)\right]
\end{align}
If we consider the subsystems length to be equal, then entanglement negativity reads 
\begin{align}
	\bar{E}^{(z)}_{N_{dis}}=\frac{3}{4}\left[\frac{10}{3N}\left(\frac{1}{2l +d}+\frac{1}{d}-\frac{2}{l  +d}\right) \right.&\left.+A_1 A_2 \left(2(l +d)^{z+1}-(2l+d)^{z+1}-d^{z+1}\right)\right.\nonumber\\&\left. +A_1 A_3 \left(2(l +d)^{\frac{z+\beta_z}{2}}-(2l +d)^{\frac{z+\beta_z}{2}}-d^{\frac{z+\beta_z}{2}}\right)\right]\label{E_N disjoint z not 2}
\end{align}
Again for arbitrary value of dynamical scaling exponent we have represented our results graphically in Fig.\eqref{Enz graph}. We would like to mention that to plot our result (in eq.\eqref{E_N disjoint z not 2}) we have to choose the value of the constants by following the inequality given in eq.\eqref{inequality z}.

\noindent
In Fig.\eqref{En different const}, we have chosen three different sets of constants and for each set we have plotted the variation of entanglement negativity with respect to the subsystem separation distance $d$. This plot is done by following the eq.\eqref{E_N disjoint z not 2} for a fixed value of dynamical scaling exponent $z=3$ and subsystem length $l=3$. The plot in green is for $c_1 $, $c_2$ and $t_{d2}$=${0.2,0.05,0.001}$, the plot in orange is for $c_1 $, $c_2$ and $t_{d2}$=${0.6,0.05,0.001}$ and the plot in purple is for $c_1 $, $c_2$ and $t_{d2}$=${0.9,0.05,0.001}$. The plot suggests that, entanglement negativity vanishes for a particular value of the separation distance for a fixed choice of subsystem length. In Fig.\eqref{En comp vs z} we have provided a graphical comparison between the excited state (for $z=3,4$ and $5$) of Lifshitz spacetime with it's ground state. The plot in green color represents the result of pure Lifshitz spacetime and The plot in purple, orange and green are for $z=3,4$ and $5$ respectively for $c_1 $, $c_2$ and $t_{d2}$=${0.6,0.05,0.001}$. We observe that for increasing value of $z$ the value of critical separation at which entanglement negativity vanishes decreases. As the entanglement negativity measures the correlation between two subsystems, for larger values of $z$ the range of correlation decreases. This plot also suggests that entanglement negativity vanishes earlier for excited state compared to the ground state of Lifshitz spacetime. \\
Again we will calculate the change in entanglement negativity due to asymptotic perturbations in the ground state Lifshitz spacetime, and the expression reads. 
\begin{align}
	\Delta \bar{E}^{(z)}_{N_{dis}}=\frac{3}{4}\left[A_1 A_2 \left(2(l +d)^{z+1}\right.\right.&\left.\left.-(2l+d)^{z+1}-d^{z+1}\right) .\right.\nonumber\\&\left. +A_1 A_3 \left(2(l +d)^{\frac{z+\beta_z}{2}}-(2l +d)^{\frac{z+\beta_z}{2}}-d^{\frac{z+\beta_z}{2}}\right)\right]\label{change EN disjoint znot2}
\end{align}
In the adjacent subsystem limit ($d \to 0$), we can also reassure that eqs.(\eqref{change EN disjoint z2} and \eqref{change EN disjoint znot2}) turns out to be eqs.(\eqref{change EN adjacent z2} and \eqref{change EN adjoint znot2}) respectively.

\noindent
For the disjoint setup, it should be emphasized from eqs.(\eqref{E_N disjoint z2},\eqref{E_N disjoint z not 2}) that the entanglement negativity does not receive any contribution from the divergent part of entanglement entropy.
\subsection{Complexity and Mutual Complexity}
In this subsection, we will compute the complexity of a subsystem of length $l$ holographicaly for the excited Lifshitz geometry. This was done in \cite{Karar:2017org} by following the "Complexity = Volume" conjecture. We know that to compute holographic complexity we have to calculate the volume of bulk region enclosed by the RT surface (parameterized by $x(r)$) and the boundary. Hence for the excited Lifshitz spacetime metric in eq.\eqref{perturbed metric} the volume is
\begin{equation}
	V=\int_{-l/2}^{l/2}dx\int_{0}^{L}dy\int_{r_{t}}^{\delta}dr~r~x(r) \sqrt{1+h_{xx}(r)+h_{yy}(r)}~.
\end{equation}
Substituting the value of $x(r)$ in the above equation, using eq.\eqref{profile} and keeping terms upto $O(h)$, we get
\begin{equation}
	V=V^{(0)}+ L\int_{r_{t}}^{\delta} dr~r~[h_{xx}(r)+h_{yy}(r)]\int_{r_{t}}^{r} du\frac{r^{2}_{t}}{u^4}\frac{1}{\sqrt{1-\frac{r^{4}_{t}}{u^{4}}}}\label{volume perturbed}~.
\end{equation}

\noindent
Hence the change in holographic complexity is defined as
\begin{equation}
	\Delta C_{A}=\frac{\Delta V}{8\pi G_{4}}\label{complexity change in volume}~.
\end{equation} 
\noindent
After substituting the values of $h_{xx}(r)$ and $h_{yy}(r)$ (in eq.\eqref{volume perturbed}) from eq.\eqref{fn z=2}, the expression of complexity for an excited state of Lifshitz spacetime , for $z=2$ reads\cite{Karar:2017org}
\begin{align}
	C^{(2)}_A = &\frac{V}{8\pi G_{4}}\nonumber\\=&\frac{\sqrt{\pi}}{8 \pi G_4 }\left[\frac{\Gamma(\frac{3}{4})L \delta^2}{\Gamma(\frac{1}{4})r^{(0)}_{t}}-\frac{\Gamma(\frac{5}{4}) L r^{(0)}_{t}}{\Gamma(\frac{3}{4})}\right] + \frac{\sqrt{\pi}\Gamma(\frac{5}{4})L}{8\pi G_{4}\Gamma(\frac{7}{4})r^{(0)3}_{t}}\left[\frac{c_{1}+c_{2}\ln r^{(0)}_{t}}{24}+\frac{3\pi +13}{288}c_{2}\right]~.
\end{align}
Now we will proceed to express the above result of complexity in terms of the subsystem length. This can be done by substituting the expression of turning point in terms of subsystem length ($l$) using eq.\eqref{length pure}. This yields
\begin{equation}
	C^{(2)}_A = \frac{\sqrt{\pi}L}{8 \pi G_4 }\left[\frac{Nl\Gamma(\frac{3}{4}) \delta^2}{2\Gamma(\frac{1}{4})}-\frac{2\Gamma(\frac{5}{4})  }{Nl\Gamma(\frac{3}{4})}\right] + \frac{N^3 \sqrt{\pi}\Gamma(\frac{5}{4})Ll^3}{64\pi G_{4}\Gamma(\frac{7}{4})}\left[\frac{c_{1}+c_{2}\ln \left(\frac{2}{Nl}\right)}{24}+\frac{3\pi +13}{288}c_{2}\right]\label{complexity for z equals 2}
\end{equation}
The change in holographic subregion complexity resulting from the perturbation in Lifshitz spacetime is defined as follows \cite{Karar:2017org}
\begin{align}
	\Delta C^{(2)}_{A}&=C^{(2)}_{A}-C^{(0)}_{A}\nonumber\\&=\frac{N^3 \sqrt{\pi}\Gamma(\frac{5}{4})Ll^3}{64\pi G_{4}\Gamma(\frac{7}{4})}\left[\frac{c_1}{24}~+\frac{3\pi +13}{288}c_{2} +\frac{c_2}{24} \ln \left(\frac{2}{Nl}\right)\right]~.
\end{align}
With this result of complexity of a subsystem of length $l$, given in  eq.\eqref{mutual complexity}, we will now compute the mutual complexity. To compute this quantity we will consider two subsystems of equal length $l$,  separated by a distance $d$. In this set up we can compute the mutual complexity by using eq.(s)(\eqref{mutual complexity},\eqref{complexity for z equals 2}). This results in the following 
\begin{align}
	\Delta \bar{C}^{(2)} &=\frac{1}{\sqrt{\pi} }\frac{\Gamma(\frac{5}{4})}{N\Gamma(\frac{3}{4})}\left[\frac{1}{2l+d}-\frac{2}{l}-\frac{1}{d}\right]\nonumber\\
	&+\frac{N^3 \Gamma(\frac{5}{4})}{16\sqrt{\pi} \Gamma(\frac{7}{4})}\left[\left(\frac{c_1}{24}+\frac{3\pi +13}{288}c_{2}\right)\left(2l^3 -(2l+d)^3 +d^3\right)\right.\nonumber\\&\left.+\frac{c_2}{24}\left(2l^3 \ln \left(\frac{2}{Nl}\right)-(2l+d)^3 \ln \left(\frac{2}{N(2l+d)}\right)-d^3 \ln \left(\frac{2}{Nd}\right)\right)\right]\label{mutual complexity z2}
\end{align}
where $\Delta \bar{C}^{(2)}=\frac{4G_4}{L}\Delta C^{(2)}$. It is to be mentioned that the above result is valid in the limit $d/l<<1$. The above result suggests that the mutual complexity has no divergence.
\begin{figure}[!h]
	\centering
	\begin{subfigure}[t]{0.5\textwidth}
		\centering
		\includegraphics[width=8.5cm]{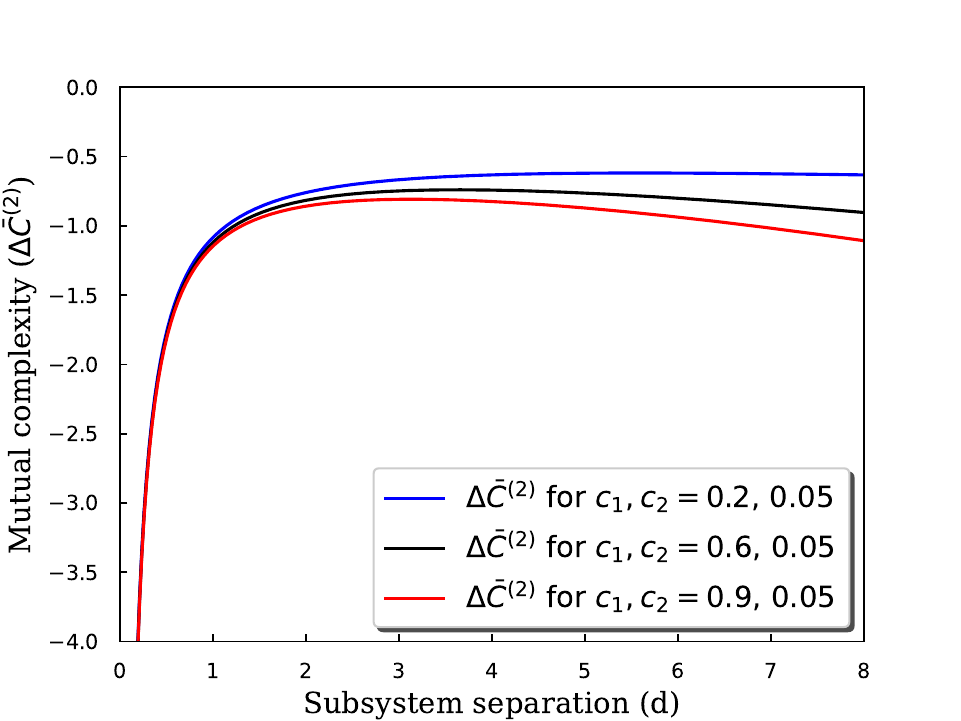}
		\caption{Different values of constants}
		\label{MC z2 different const}
	\end{subfigure}%
	~ 
	\begin{subfigure}[t]{0.5\textwidth}
		\centering
		\includegraphics[width=8.5cm]{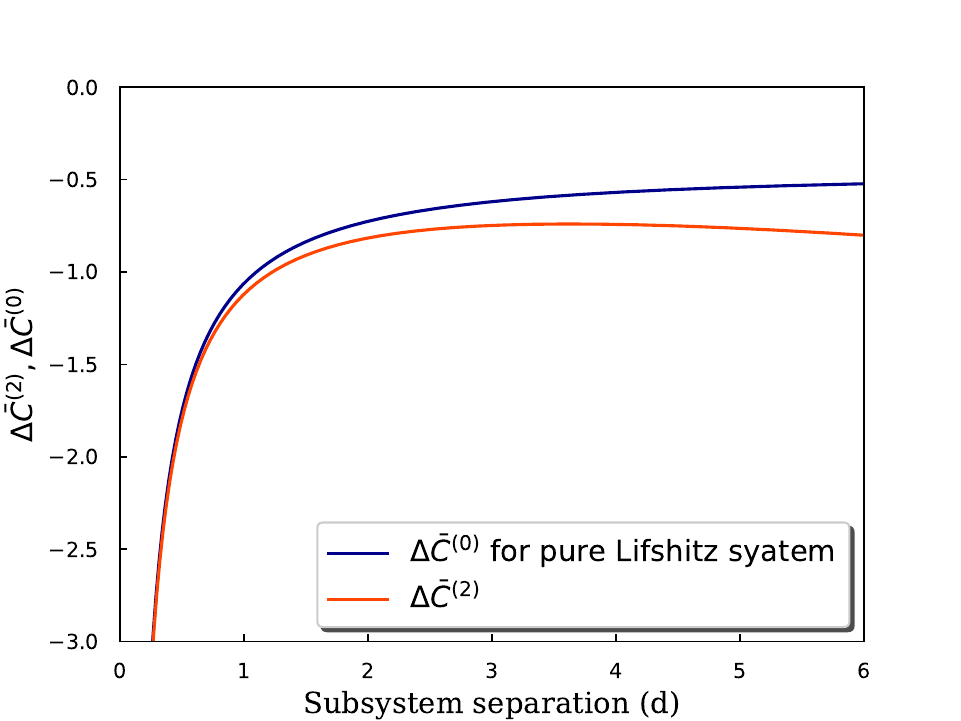}
		\caption{Comparison with the pure system}
		\label{MC z2 pure compare}
	\end{subfigure}
	\caption{The above plots shows the variation of mutual complexity for two disjoint subsystems (with length $l=3$) with respect to their separation distance. In the left panel of the Figure, we have shown the variation of mutual complexity with respect to the subsystem separation ($d$) for a fixed subsystem length $l=3$. The plots are done for different values of the constants. In particular, the plot in blue is for $c_1 $, $c_2$=${0.2,0.05}$, the plot in black is for $c_1 $, $c_2$=${0.6,0.05}$, and the plot in red is for $c_1 $, $c_2$=${0.9,0.05}$. In the right panel of the Figure, we have graphically shown a comparison between the results of pure and excited state (for $c_1 $, $c_2$=${0.6,0.05}$) Lifshitz spacetime. The plot in blue represents the result of mutual complexity for the pure Lifshitz spacetime and the plot in red is for excited state Lifshitz spacetime with $c_1 $, $c_2$=${0.6,0.05}$ and $z=2$.}
	\label{Mc z2,pure}
\end{figure}

\begin{figure}[!h]
	\centering
	\begin{subfigure}[t]{0.5\textwidth}
		\centering
		\includegraphics[width=8.5cm]{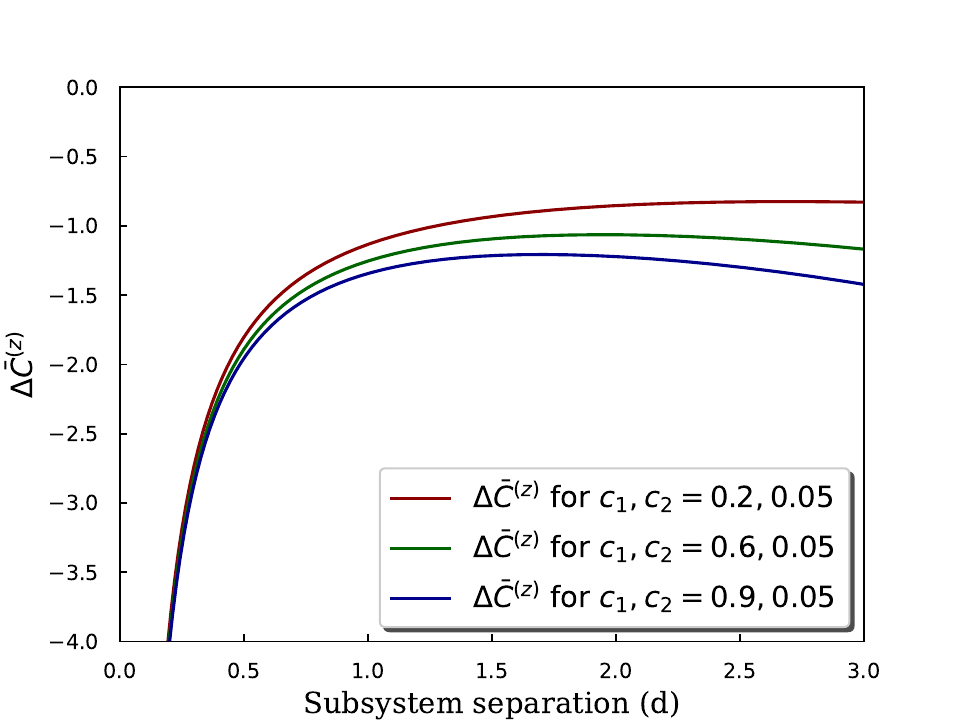}
		\caption{Plots for different values of the constants for fixed dynamical exponent ($z=3$)}
		\label{mcz different constants}
	\end{subfigure}%
	~ 
	\begin{subfigure}[t]{0.5\textwidth}
		\centering
		\includegraphics[width=8.5cm]{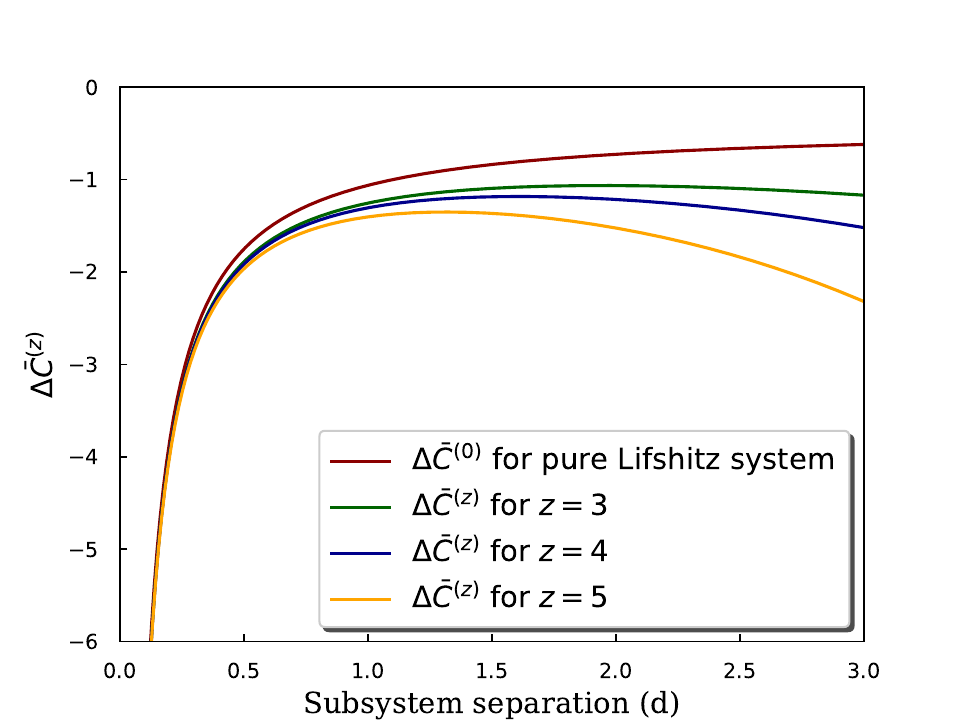}
		\caption{Comparison between pure and excited system}
		\label{mcz pure vs}
	\end{subfigure}\\[1ex]
	\caption{The above plots show the variation of mutual complexity with respect to the subsystem separation. In Fig.(8a), we have shown the variation of mutual complexity with respect to the subsystem separation for a fixed value of subsystem length $l=3$ and dynamical exponent $z=3$. The plot in brown is for $c_1 $, $c_2$=${0.2,0.05}$, the plot in green is for $c_1 $, $c_2$=${0.6,0.05}$ and the plot in blue is for $c_1 $, $c_2$=${0.9,0.05}$. In Fig.(8b), we have shown a graphical comparison between the results of mutual complexity for pure and excited state ($z=3,4$ and $5$) of Lifshitz spacetime. The plot in brown refers to the pure state mutual complexity of Lifshitz spacetime and the graph in color green, blue and orange are for $z=3,4$ and $5$ respectively with $c_1 $, $c_2$=${0.6,0.05}$.}
\end{figure}

\noindent
We will now plot the result of mutual complexity (given in eq.\eqref{mutual complexity z2}) with respect to subsystem separation distance $d$. In order to obtain the plots, we have chosen the constants from the bound as mentioned in eq.\eqref{inequality const}. In Fig.\eqref{MC z2 different const}, we have chosen three different sets of constants and for each set we have shown the variation of mutual complexity with respect to the subsystem separation. The plot in blue is for $c_1 $, $c_2$=${0.2,0.05}$, the plot in black is for $c_1 $, $c_2$=${0.6,0.05}$, and the plot in red is for $c_1 $, $c_2$=${0.9,0.05}$. It is clear from the plot that for each set of constants the mutual complexity is always negative, hence it is superadditive in nature. In Fig.\eqref{MC z2 pure compare}, we have graphically compared the variation of mutual complexity for excited state (for $z=2$) and ground state of Lifshitz spacetime. From the plot it is clear that the mutual complexity corresponding to pure Lifshitz spacetime is less negative compared to the excited state for $c_1 $, $c_2$=${0.6,0.05}$. The plot in blue represents the result of mutual complexity for the pure Lifshitz spacetime and the plot in red is for excited state Lifshitz spacetime with $c_1 $, $c_2$=${0.6,0.05}$ and $z=2$. 
The change in mutual complexity for a perturbation applied to the ground state of Lifshitz spacetime is defined by
\begin{align}
	\delta C^{(2)}=\Delta C^{(2)} -\Delta C^{(0)}=&\frac{N^3 \sqrt{\pi}\Gamma(\frac{5}{4})L}{64\pi G_{4}\Gamma(\frac{7}{4})}\left[\left(\frac{c_1}{24}+\frac{3\pi +13}{288}c_{2}\right)\left(2l^3 -(2l+d)^3 +d^3\right)\right.\nonumber\\&\left.+\frac{c_2}{24}\left(2l^3 \ln \left(\frac{2}{Nl}\right)-(2l+d)^3 \ln \left(\frac{2}{N(2l+d)}\right)-d^3 \ln \left(\frac{2}{Nd}\right)\right)\right]~.\label{change in MC z2}
\end{align}
In order to compute the complexity of subsystem of length $l$ for arbitrary value of dynamical scaling exponent ($z$), we have to perform the integral as given in eq.\eqref{volume perturbed}, to do so we will substitute the expression of $h_{xx}(r)$ and $h_{yy}(r)$ from the eq.(s)(\eqref{pertutbations},\eqref{fn z not 2}), that gives
\begin{align}
	C_{A}^{(z)}&=\frac{\sqrt{\pi}}{8 \pi G_4 }\left[\frac{\Gamma(\frac{3}{4})L \delta^2}{\Gamma(\frac{1}{4})r^{(0)}_{t}}-\frac{\Gamma(\frac{5}{4}) L r^{(0)}_{t}}{\Gamma(\frac{3}{4})}\right]\nonumber\\
	&+\frac{\sqrt{\pi}L}{8\pi G_{4}r^{(0)(z+1)}_{t}}\left[\frac{c_{1}\Gamma \left(\frac{z+3}{4}\right)}{z(z+2)\Gamma \left(\frac{z+5}{4}\right)}+r^{(0)\frac{1}{2}(z+2-\beta_{z})}_{t}\frac{2c_{2}(4+\beta_{z} -3z)\Gamma \left(\frac{z+\beta_{z}+4}{8}\right)}{(z+\beta_{z}+2)(z+\beta_{z}-2)\Gamma \left(\frac{z+\beta_{z}+8}{8}\right)}\right]~.
\end{align}
Now we will rewrite the expression of complexity in terms of subsystem length. To do this we have to substitute the expression given in eq.\eqref{length pure} in the above result. This yields
\begin{align}
	C_{A}^{(z)}=&\frac{\sqrt{\pi}L}{8 \pi G_4 }\left[\frac{Nl\Gamma(\frac{3}{4}) \delta^2}{2\Gamma(\frac{1}{4})}-\frac{2\Gamma(\frac{5}{4})  }{Nl\Gamma(\frac{3}{4})}\right]\nonumber\\
	&+\frac{\sqrt{\pi}(Nl)^{z+1}L}{8\pi G_{4}2^{z+1}}\left[\frac{c_{1}\Gamma \left(\frac{z+3}{4}\right)}{z(z+2)\Gamma \left(\frac{z+5}{4}\right)}+\left(\frac{2}{Nl}\right)^{\frac{1}{2}(z+2-\beta_{z})}\frac{2c_{2}(4+\beta_{z} -3z)\Gamma \left(\frac{z+\beta_{z}+4}{8}\right)}{(z+\beta_{z}+2)(z+\beta_{z}-2)\Gamma \left(\frac{z+\beta_{z}+8}{8}\right)}\right]~.\label{complexity znot2 total}
\end{align}
\noindent
The change in complexity due to the metric perturbation is given by 
\begin{equation}
	\Delta C_{A}^{(z)}=\frac{\sqrt{\pi}(Nl)^{z+1}L}{8\pi G_{4}2^{z+1}}\left[\frac{c_{1}\Gamma \left(\frac{z+3}{4}\right)}{z(z+2)\Gamma \left(\frac{z+5}{4}\right)}+\left(\frac{2}{Nl}\right)^{\frac{1}{2}(z+2-\beta_{z})}\frac{2c_{2}(4+\beta_{z} -3z)\Gamma \left(\frac{z+\beta_{z}+4}{8}\right)}{(z+\beta_{z}+2)(z+\beta_{z}-2)\Gamma \left(\frac{z+\beta_{z}+8}{8}\right)}\right]\label{complexity znot2}
\end{equation}
Now we will proceed to compute the mutual complexity between two subsystems of equal length $l$ and separation distance $d$. To compute this we will follow the similar procedure as described for $z=2$ case. The mutual complexity can be computed by using eq.(s)(\eqref{mutual complexity},\eqref{complexity znot2 total}). This results in the following
\begin{align}
	\Delta \bar{C}^{(z)}&=\frac{1}{\sqrt{\pi}}\frac{\Gamma(\frac{5}{4})}{N\Gamma(\frac{3}{4})}\left[\frac{1}{2l+d}-\frac{2}{l}-\frac{1}{d}\right]\nonumber\\&+\frac{N^{z+1}}{2\sqrt{\pi} 2^{z+1}}\left[\frac{c_{1}\Gamma \left(\frac{z+3}{4}\right)}{z(z+2)\Gamma \left(\frac{z+5}{4}\right)}\left\{2l^{z+1}-(2l+d)^{z+1}+d^{z+1}\right\}+\right.\nonumber\\&\left.\left(\frac{2}{N}\right)^{\frac{1}{2}(z+2-\beta_{z})}\frac{2c_{2}(4+\beta_{z}-3z)}{(z+\beta_{z}+2)(z+\beta_{z}-2)}\frac{\Gamma\left(\frac{z+\beta_{z}+4}{8}\right)}{\Gamma\left(\frac{z+\beta_{z}+8}{8}\right)}\left\{2l^{\frac{z+\beta_{z}}{2}}-(2l+d)^{\frac{z+\beta_{z}}{2}}+d^{\frac{z+\beta_{z}}{2}}\right\}\right]\label{total MC znot2}
\end{align}
where $\Delta \bar{C}^{(z)}=\frac{4G_4}{L}\Delta C^{(z)}$.\\
We will now proceed to show that the variation of the $ \Delta C^{(z)}$ with respect to subsystem separation $d$ graphically. To do so we have chosen the value of constants from the inequality given in eq.\eqref{inequality z}.
In Fig.\eqref{mcz different constants}, we have again chosen three different sets of constants and for each set we have plotted the mutual complexity with respect to the subsystem separation ($d$). The plot in brown is for $c_1 $, $c_2$=${0.2,0.05}$, the plot in green is for $c_1 $, $c_2$=${0.6,0.05}$, and the plot in blue is for $c_1 $, $c_2$=${0.9,0.05}$. It is clear that for each set of $c_1 ,c_2$ and $t_{d2}$, the mutual complexity ($\Delta C^{(z)}$) is negative, hence it is superadditive in nature.

\noindent
In Fig.\eqref{mcz pure vs}, we have shown a comparison between the mutual complexity of ground and an excited state (for $z=3,4,5$ and $c_1 ,c_2 =0.6,0.05$) of Lifshitz spacetime. The graph in brown refers to the pure state mutual complexity of Lifshitz spacetime and the graph in color green, blue and orange are for $z=3,4$ and $5$ respectively with $c_1 $, $c_2$=${0.6,0.05}$.
This figure also tells us that for the excited state mutual complexity is always more negative than that of the ground state.  Hence, for each value of dynamical scaling exponent the mutual complexity of the excited Lifshitz spacetime is superadditive in nature.
\vskip 0.2cm
\noindent 
Just like the previous scenarios, we will also compute the change in mutual complexity, and that is defined as follows
\begin{equation}
	\delta C^{(z)}=\Delta C^{(z)}-\Delta C^{(0)}~.
\end{equation}
Hence, we have
\begin{align}
	\delta C^{(z)}&=\frac{\sqrt{\pi} L N^{z+1}}{8\pi G_{4}2^{z+1}}\left[\frac{c_{1}\Gamma \left(\frac{z+3}{4}\right)}{z(z+2)\Gamma \left(\frac{z+5}{4}\right)}\left\{2l^{z+1}-(2l+d)^{z+1}+d^{z+1}\right\}+\right.\nonumber\\&\left.\left(\frac{2}{N}\right)^{\frac{1}{2}(z+2-\beta_{z})}\frac{2c_{2}(4+\beta_{z}-3z)}{(z+\beta_{z}+2)(z+\beta_{z}-2)}\frac{\Gamma\left(\frac{z+\beta_{z}+4}{8}\right)}{\Gamma\left(\frac{z+\beta_{z}+8}{8}\right)}\left\{2l^{\frac{z+\beta_{z}}{2}}-(2l+d)^{\frac{z+\beta_{z}}{2}}+d^{\frac{z+\beta_{z}}{2}}\right\}\right]~.\label{change in MC znot2}
\end{align}
\section{Thermodynamic relations between different information theoretic \\quantities}\label{section 4}
In recent years thermodynamics of quantum information theoretic quantities have been studied extensively \cite{Allahbakhshi:2013rda,Bhattacharya:2012mi,Mukohyama:1997ww,Wong:2013gua}. The thermodynamic relation for HEE and HSC for perturbed Lifshitz geometry have already been computed in \cite{Chakraborty:2014lfa,Karar:2017org} respectively. In this section, we will establish relationships similar to the first law of entanglement thermodynamics among various mixed state information theoretic measures (for entanglement wedge cross section, entanglement negativity and mutual complexity) for an excited state in Lifshitz spacetime. To do so we first need to know how the energy, entanglement pressure and entanglement chemical potential change due to the metric perturbation. For both the values of dynamical scaling exponent $z$. These quantities can be obtained by calculating the vacuum expectation values(VEV) of the stress energy tensor, given in eqs.(\eqref{stess tensor}). Following \cite{Chakraborty:2014lfa}, we can calculate the VEV of the stress tensor given in eq.\eqref{stess tensor}. Before proceeding further we need to mention that, the stress tensor has contributions from the massive gauge field. Hence, 
we also need to mention about another function $s_{0}(r)$ which can be obtained by varying the linearized action with respect to massive gauge field and the functional form reads\cite{Ross:2009ar}\footnote{The quantity $s_{0}(r)$ is itself not conserved and dual to the gauge field $A^{0}=r^{-z}A_{t}=\alpha$\cite{Ross:2009ar,Chakraborty:2014lfa}.}
\begin{equation}
	s_{0}(r)=\alpha r^{z+2}\left[z j(r)+r \partial_r (\frac{1}{2}f(r)+j(r))\right]~.
\end{equation}
The expression of $s_{0}(r)$ will be useful when we will calculate the change in entanglement chemical potential later.
After substituting the form of the perturbation functions ($z=2$ scenario) in eq.\eqref{stess tensor} from eq.\eqref{fn z=2} one obtains\cite{Ross:2009ar,Chakraborty:2014lfa}
\begin{align}
	&\left<T_{tt}\right>=\frac{1}{16 \pi G_4}\frac{4 c_2}{3}\nonumber\\
	&\left<T_{xx}\right>=\frac{1}{16 \pi G_4}\left(\frac{4 c_2}{3}+4t_{d2}\right)~.\label{stress tensor z2}
\end{align}
Now for $z=2$ the VEV for the function $s_{0}(r)$ is given by \cite{Chakraborty:2014lfa}
\begin{equation}
	\big<s_{0}(r)\big>=\frac{4}{3}\left(c_1 +c_2 \ln r\right)~.\label{s0 r}
\end{equation}
One can see that the quantity $\big<s_{0}(r)\big>/16\pi G_{4}$ has the same scaling dimension as the chemical potential. Keeping this in mind one can express the chemical potential of excited state in terms of $s_{0}(r)$. This results
\begin{equation}
	\Delta \mu = \frac{\left<s_{0}(r_{t})\right>}{16\pi G_{4}}~.\label{change in chemical potential}
\end{equation}
Similarly for $z\neq 2$, after substituting the perturbation functions from eq.\eqref{fn z not 2} into eqs.(\eqref{stess tensor},\eqref{change in chemical potential}), one can get the following expressions \cite{Ross:2009ar,Chakraborty:2014lfa}

\begin{align}
	&\left<T_{tt}\right>=\frac{1}{4 \pi G_4}\left(\frac{z-2}{z}\right)c_1\nonumber\\
	&\Delta \mu=\frac{\left<s_{0}(r)\right>}{16 \pi G_{4}}=\frac{1}{16 \pi G_4}\frac{\alpha}{(z-1)}\left[4c_1+c_2 z(4+\beta_z-3z)r^{(0)\frac{1}{2}(z+2-\beta_z)}_{t}\right]\nonumber\\
	&\left<T_{xx}\right>=\frac{1}{16 \pi G_4}[2(z-2)c_1 + (z+2)t_{d2}]~.\label{stress tensor znot2}
\end{align}
Inverting the above two expressions leads to
\begin{align}
	&c_1 =4 \pi G_4 \frac{z}{z-2}\left<T_{tt}\right>\nonumber\\
	&c_2 =16 \pi G_4 \left[\frac{(z-1)(z-2)\Delta \mu - \alpha z \left<T_{tt}\right>}{\alpha z (z-2)(4+\beta_z -3z)r^{(0)\frac{1}{2}(z+2-\beta_z)}_{t}}\right]\nonumber\\
	&t_{d2}=16 \pi G_4 \left[\frac{2\left<T_{xx}\right>-z\left<T_{tt}\right>}{2(z+2)}\right]~.\label{c_1 c_2 znot2}
\end{align}

\noindent
For $z=2$, the total energy, entanglement pressure, entanglement chemical potential and charge for the excited state is given by\cite{Chakraborty:2014lfa}
\begin{eqnarray}
	&\Delta E=\int_{0}^{L}dy \int_{-l/2}^{l/2}dx \left<T_{tt}\right>=Ll\left<T_{tt}\right>\nonumber\\
	&\Delta P_{x}=\left<T_{xx}\right>\nonumber\\
	&\Delta \mu =\frac{1}{12\pi G_{4}}\left(c_1 +c_2 \ln r^{(0)}_{t}\right)\nonumber\\
	&Q=m^2 \alpha Ll=4Ll~.\label{thermo quantities }
\end{eqnarray}
As the system is not in thermal equilibrium, it is important to clarify that the pressure ($P_{x}$) should not be confused with the pressure in thermodynamics. Instead, we will refer to it as entanglement pressure.

\noindent
Following \cite{Allahbakhshi:2013rda,Bhattacharya:2012mi,Mukohyama:1997ww,Wong:2013gua}, entanglement temperature is defined in terms of length of the strip like subsystems. The entanglement temperature is inversely related to the size of the entangling region, and the specific constant of proportionality is determined by the shape of the entangling region. For $z=2$, the entanglement temperature can be expressed as
\begin{equation}
	T_{ent}=\frac{24r^{2}_{t}}{\pi}\frac{25}{(324-30\pi)}=\frac{192}{\pi N^2 l^2}\frac{25}{(324-30\pi)}~.\label{ent temp z2}
\end{equation} 

\noindent
The first law of entanglement thermodynamics for excited state of Lifshitz spacetime is\cite{Chakraborty:2014lfa}
\begin{equation}
	\Delta E=T_{ent}\Delta S +\frac{10}{(54-5\pi)}\Delta P_{x}V -\frac{5}{(54-5\pi)}\Delta \mu Q~.\label{Energy z2}
\end{equation}
For $z\neq 2$, the entanglement temperature \cite{Allahbakhshi:2013rda,Bhattacharya:2012mi,Mukohyama:1997ww,Wong:2013gua} is defined as
\begin{equation}
	T_{ent}=\frac{2r^{(0)z}_{t}\Gamma(\frac{3}{4})}{\pi \Gamma(\frac{1}{4})\tilde{A_1}}\label{ent temp znot2}
\end{equation}
When $z \neq 2$, the first law of entanglement thermodynamics is formulated as \cite{Chakraborty:2014lfa}
\begin{equation}
	\Delta E=T_{ent}\Delta S+\frac{\tilde{A_2}}{\tilde{A_1}}\Delta P_{x}V-\frac{\tilde{A_3}}{\tilde{A_1}}\Delta \mu Q \label{Energy znot2}
\end{equation}
where 
\begin{align}
	&\tilde{A_1} =\frac{z^2 }{(z+3)(z^2 -4)}\frac{\Gamma \left(\frac{z+1}{4}\right)}{\Gamma \left(\frac{z+3}{4}\right)}-\frac{2}{(z-2)(4+z+\beta_z)}\frac{\Gamma \left(\frac{z+\beta_z}{8}\right)}{\Gamma \left(\frac{z+4+\beta_z}{8}\right)}\\
	&\tilde{A_2} =\frac{\Gamma \left(\frac{z+1}{4}\right)}{(z+2)(z+3)\Gamma \left(\frac{z+3}{4}\right)},~~~
	\tilde{A_3} =\frac{\Gamma \left(\frac{z+\beta_z}{8}\right)}{2z(4+z+\beta_z)\Gamma \left(\frac{z+4+\beta_z}{8}\right)}~.
\end{align}
where $Q=m^2 \alpha Ll=\sqrt{8z(z-1)}$.

\noindent
With all this quantities in hand, we shall now proceed to relate the change in EWCS, entanglement negativity, mutual complexity with change in energy density ($\Delta E$), entanglement pressure ($\Delta P_{x}$) and  entanglement chemical potential ($\Delta \mu$).
\subsection{Thermodynamic relation for EWCS}\label{4.1}
In order to define a first law like relation of entanglement thermodynamics for the change in EWCS ($z=2$ scenario), we will take the limit $d$ approaching to zero in eq.\eqref{change EW z2}, which leads to\footnote{We will follow the procedure mentioned in \cite{ChowdhuryRoy:2022dgo} to derive the first law like relation of entanglement thermodynamics for the mixed state information theoretic quantities.}
\begin{equation}
	\Delta \bar{E}^{(2)}_{W}=\frac{N^3 l^3}{6}\left[\left(\frac{4c_1 +5c_2}{24}-t_{d2}+\frac{c_2}{18}\right)+\frac{c_2}{6}\ln\left(\frac{1}{Nl}\right)\right]~.
\end{equation}
Substituting the form of $c_1$, $c_2$ and $t_{d2}$ from eq.(s)(\eqref{stress tensor z2},\eqref{s0 r}) in the above expression of change in EWCS and using eqs.(\eqref{thermo quantities },\eqref{ent temp z2}), we get
\begin{equation}
	\Delta E^{(2)}_{W}=\frac{1}{T_{ent}}\left[B_1 \Delta \mu Q +B_2 \Delta P_{x}V +B_3 \Delta E\right]~.
\end{equation}

\noindent
Rewriting the above equation leads to the expression of change in energy due to perturbation, and that reads
\begin{equation}
	\Delta E=\frac{1}{B_3}\left[T_{ent}\Delta E_{W}-B_1 \Delta \mu Q-B_2 \Delta P_{x}V\right]\label{EW energy z2}
\end{equation} 
where 
\begin{align}
	&B_1 =\frac{50N}{(162-15\pi)},~~~B_2 =-\frac{400N}{(162-15\pi)}=-8B_1,\\&B_3 =4\left(\frac{43}{72}-\frac{\ln 2}{6}\right)\frac{50N}{(162-15\pi)}=4\left(\frac{43}{72}-\frac{\ln 2}{6}\right)B_1~.
\end{align}
Equating eqs.(\eqref{EW energy z2},\eqref{Energy z2}), we get the following relation for the change in EWCS
\begin{equation}
	\Delta E^{(2)}_{W}=B_{3}\Delta S+\left(\frac{10B_{3}}{54-5\pi}+B_1\right)\frac{\Delta P_{x}V}{T_{ent}}+\left(B_2 -\frac{5B_3}{54-5\pi}\right)\frac{\Delta \mu Q}{T_{ent}}~.\label{thermo EW z2}
\end{equation}
Following the same procedure as $z=2$ scenario, we will now proceed to compute a thermodynamic relation for the change in EWCS due to metric perturbation for arbitrary value of the dynamical scaling exponent. Taking the limit $d\to 0$ in eq.\eqref{change in EW for z} and using the expressions of $c_1$, $c_2$, $t_{d2}$ and $T_{ent}$ from eq.(s)(\eqref{c_1 c_2 znot2},\eqref{ent temp znot2}) leads to the following result
\begin{equation}
	\Delta E^{(z)}_{W}=\frac{1}{T_{ent}}\left[F_{1}\Delta E -F_{2}\Delta P_{x}V +F_{3}\Delta \mu Q\right]~.
\end{equation}
Now to obtain a first law like relation of entanglement thermodynamics for the change in EWCS, we have to substitute the expression of $\Delta E$ from eq.\eqref{Energy znot2} in the above equation. This yields
\begin{equation}
	\Delta E^{(z)}_{W}=F_{1}\Delta S +\left(\frac{F_{1}\tilde{A_2}}{\tilde{A_1}}-F_{2}\right)\frac{\Delta P_{x}V}{T_{ent}}+\left(\frac{F_{1}\tilde{A_3}}{\tilde{A_1}}+F_{3}\right)\frac{\Delta \mu Q}{T_{ent}}\label{thermo EW znot2}
\end{equation}
where
\begin{align}
	&F_{1}=\frac{2^{z+1}N\Gamma(\frac{3}{4})}{\Gamma(\frac{1}{4})\tilde{A_1}}\left[\frac{z(z-1)}{(z+1)(z^2 -4)}-\frac{2^{-\frac{1}{2}(z-\beta_z -4)}}{(z+\beta_z)(z-2)(z+2+\beta_z)}\right]\nonumber\\
	&F_{2}=\frac{2^{z+2}N\Gamma(\frac{3}{4})}{(z+1)(z+2)\Gamma(\frac{1}{4})\tilde{A_1}}\nonumber\\
	&F_{3}=\frac{2^{z+1}N\Gamma(\frac{3}{4})}{\Gamma(\frac{1}{4})\tilde{A_1}}\left[\frac{(z-1)2^{-\frac{1}{2}(z-\beta_z -4)}}{m^2 \alpha^2 z(z+\beta_z)(z+2+\beta_z) }\right]~.
\end{align}
The change in EWCS in the eqs.(\eqref{thermo EW z2},\eqref{thermo EW znot2}) represents a first law like relation of entanglement thermodynamics for $z=2$ and $z \neq 2$ respectively. Both the thermodynamic relation has similar forms except the constant terms. One should note that the term $\Delta P_{x}$ appears in eqs.(\eqref{thermo EW z2},\eqref{thermo EW znot2}) due to the presence of the $t_{d2}$ term in the expression of EWCS, which arises from the term $h_{yy}(r)$ inside the integral. 
\subsection{Thermodynamic relation for entanglement negativity} \label{4.2}
In this subsection we will establish a thermodynamic relation for the change in entanglement negativity for both the values of the dynamical scaling exponent. After substituting the values of $c_1$, $c_2$ and $t_{d2}$ from eq.\eqref{stress tensor z2} in eq.\eqref{change EN adjacent z2} and using eqs.(\eqref{thermo quantities },\eqref{ent temp z2} and \eqref{change in chemical potential}), we get
\begin{equation}
	\Delta E^{(2)}_{N}=\frac{1}{T_{ent}}\left[-D_{1}\Delta \mu Q +D_{2}\Delta P_{x}V + D_{3}\Delta E\right]~.
\end{equation}
Now in the above equation, we will substitute the value of $\Delta E$ from \eqref{Energy z2}, and after a little bit of rearrangement the first law like relation for change in $E_{N}$ ($z=2$ scenario) reads
\begin{equation}
	\Delta E^{(2)}_{N}=D_{3}\Delta S +\left(\frac{10D_{3}}{54-5\pi}+D_{2}\right)\frac{\Delta P_{x}V}{T_{ent}}-\left(\frac{5D_{3}}{54-5\pi}+D_{1}\right)\frac{\Delta \mu Q}{T_{ent}}
\end{equation}
where 
\begin{equation}
	D_{1}=-\frac{90}{54-5\pi},~~D_{2}=\frac{180}{54-5\pi}~and~~D_{3}=\frac{360}{54-5\pi}\left(\frac{\ln 16}{3}-\frac{27}{10}+\frac{\pi}{4}\right)~.
\end{equation}
For arbitrary value of dynamical exponent ($z$), we can also express the change in negativity with change in energy, entanglement pressure and entanglement chemical potential. Using eq.(s)(\eqref{c_1 c_2 znot2},\eqref{ent temp znot2}), if we substitute the expressions of $c_1$, $c_2$, $t_{d2}$ and $T_{ent}$ in the expression of change in entanglement negativity (eq.\eqref{change EN adjoint znot2}) for arbitrary $z$, we get the following expression 
\begin{equation}
	\Delta E^{(z)}_{N}=\frac{1}{T_{ent}}\left[D_{4}\Delta E -D_{5}\Delta P_{x}V+D_{6}\Delta \mu Q\right]~.
\end{equation}
After substituting the expression of $\Delta E$ from eq.\eqref{Energy znot2}, the above equation becomes
\begin{equation}
	\Delta E^{(z)}_{N}=D_{4}\Delta S +\left(\frac{D_{4}\tilde{A_2}}{\tilde{A_1}}-D_{5}\right)\frac{\Delta P_{x}V}{T_{ent}}+\left(D_{6}-\frac{D_{4}\tilde{A_3}}{\tilde{A_1}}\right)\frac{\Delta \mu Q}{T_{ent}}
\end{equation}
where
\begin{align}
	&D_{4}=\frac{3}{2\tilde{A_1}}\left[\left(\frac{z^2}{z^2 -4}\right)\frac{\Gamma \left(\frac{z+1}{4}\right)}{\Gamma \left(\frac{z+3}{4}\right)}\frac{(1-2^{z+1})}{(z+3)}-\frac{\Gamma \left(\frac{z+\beta_z}{8}\right)}{\Gamma \left(\frac{z+4+\beta_z}{8}\right)}\frac{(1-2^{\frac{z+\beta_z -2}{2}})}{(z-2)(4+z+\beta_z)}\right]\nonumber\\
	&D_{5}=\frac{3}{2\tilde{A_1}}\left[\frac{\Gamma \left(\frac{z+1}{4}\right)}{\Gamma \left(\frac{z+3}{4}\right)}\frac{(1-2^{z+1})}{(z+1)(z+2)}\right]\nonumber\\
	&D_{6}=\frac{3}{2\tilde{A_1}}\frac{\Gamma \left(\frac{z+\beta_z}{8}\right)}{\Gamma \left(\frac{z+4+\beta_z}{8}\right)}\frac{(z-1)(1-2^{\frac{z+\beta_z -2}{2}})}{(4+z+\beta_z)m^2 \alpha^2 z}~.
\end{align}
We can see that the thermodynamic relation for the entanglement negativity for both the values dynamical exponent has similar mathematical form except the constants. The thermodynamic relations contains $\Delta P_{x}$, which is due to the presence of $h_{xx}(r)+\left(\frac{r^{(0)}_{t}}{r}\right)^2 h_{yy}(r)$ term in the expression of entanglement entropy.
\subsection{Thermodynamic relation for mutual complexity}
In this subsection we will derive a first law like relation for entanglement thermodynamics for the change in mutual complexity (for both the values of $z$). We will follow the same procedure as discussed in the previous subsections (\eqref{4.1},\eqref{4.2}). In order to do so we will again take the limit $d \to 0$ in eq.\eqref{change in MC z2}, this will lead to the following relation 
\begin{equation}
	\delta C^{(2)} =\frac{N^3 \sqrt{\pi}\Gamma\left(\frac{5}{4}\right)Ll^3}{64\pi G_{4}\Gamma\left(\frac{7}{4}\right)}\left[-\left(\frac{c_1}{4}+\frac{3\pi + 13}{48}c_2\right) + \frac{c_2}{12}\ln(2N^3 l^3)\right]~.
\end{equation}
Now in the above equation, substituting the values of $c_1$ and $c_2$ from eq.\eqref{stress tensor z2} and using eqs.(\eqref{thermo quantities },\eqref{ent temp z2}) gives us 
\begin{equation}
	\delta C^{(2)} =\frac{1}{T_{ent}}\left[-K_{1}\Delta \mu Q + K_{2}\Delta E\right]~.
\end{equation}
To write an entanglement thermodynamic relation for the change in mutual complexity (for $z=2$ scenario), we will substitute the expression of $\Delta E$ from eq.\eqref{Energy z2} in the above relation, after simplification that reads
\begin{equation}
	\delta C^{(2)} =K_{2}\Delta S +\frac{10K_{2}}{54-5\pi}\frac{\Delta P_{x}V}{T_{ent}}-\left(\frac{5K_{2}}{54-5\pi}+K_{1}\right)\frac{\Delta \mu Q}{T_{ent}}
\end{equation}
where the constants $K_{1}$ and $K_{2}$ are respectively
\begin{align}
	&K_{1}=\frac{75N\sqrt{\pi}\Gamma\left(\frac{5}{4}\right)}{16\Gamma \left(\frac{7}{4}\right)\pi (54-5\pi)}\nonumber\\
	&K_{2}=\frac{75N\sqrt{\pi}\Gamma\left(\frac{5}{4}\right)}{\Gamma \left(\frac{7}{4}\right)\pi (54-5\pi)}\left(\frac{\ln 2}{3}-\frac{3\pi +13}{48}\right)~.
\end{align}
Now we will proceed to establish a thermodynamic relation for the change in mutual complexity for arbitrary value of the dynamical scaling exponent. As the subsystem separation approaches zero, eq.\eqref{change in MC znot2} reduces to the following form for the change in mutual complexity
\begin{equation}
	\delta C^{(z)}=\resizebox{.91\hsize}{0.034\vsize}{$\frac{\sqrt{\pi}LN^{z+1}}{8\pi G_{4}2^{z+1}}\left[\frac{2c_{1}\Gamma\left(\frac{z+3}{4}\right)(1-2^{z})l^{z+1}}{z(z+2)\Gamma\left(\frac{z+5}{4}\right)}+\frac{2^{\frac{1}{2}(z+2-\beta_{z})}4c_{2}(4+\beta_{z}-3z)\Gamma\left(\frac{z+\beta_{z}+4}{8}\right)(1-2^{\frac{z+\beta_{z}-2}{2}}) }{N^{\frac{1}{2}(z+2-\beta_{z})}(z+\beta_{z}+2)(z+\beta_{z}-2)\Gamma\left(\frac{z+\beta_{z}+8}{8}\right)}l^{\frac{z+\beta_{z}}{2}}\right]$}	
\end{equation}

\noindent
If we substitute the values of $c_1$, $c_2$ from eq.\eqref{stress tensor znot2} and use eq.(\eqref{thermo quantities }), the above expression of mutual complexity becomes
\begin{equation}
	\delta C^{(z)}=\frac{1}{T_{ent}}\left[K_{3}\Delta E + K_{4}\Delta \mu Q\right]~.
\end{equation}
where
\begin{align}
	&K_{3}=\frac{1}{\pi \tilde{A_1}}\left[\frac{\Gamma\left(\frac{z+3}{4}\right)(1-2^{z})}{(z^{2}-4)\Gamma\left(\frac{z+5}{4}\right)}-\frac{8\Gamma\left(\frac{z+\beta_{z}+4}{8}\right)(1-2^{\frac{z+\beta_{z}-2}{2}})}{(z-2)(z+\beta_{z}+2)(z+\beta_{z}-2)\Gamma\left(\frac{z+\beta_{z}+8}{8}\right)}\right]\nonumber\\
	&K_{4}=\frac{2\Gamma\left(\frac{z+\beta_{z}+4}{8}\right)2^{\frac{z+\beta_{z}-2}{2}}}{\pi \tilde{A_1}z(z-2)(z+\beta_{z}+2)(z+\beta_{z}-2)\Gamma\left(\frac{z+\beta_{z}+8}{8}\right)}~.
\end{align}
To establish a relationship between mutual complexity (for arbitrary value of $z$), entanglement pressure and chemical potential, using eq.\eqref{Energy znot2}, we will put the expression of $\Delta E$ in the above equation of $\delta C$, and this reads
\begin{equation}
	\delta C^{(z)}=\frac{1}{T_{ent}}\left[K_{3}\Delta S +\frac{K_{3}\tilde{A_2}}{\tilde{A_1}}\frac{\Delta P_{x}V}{T_{ent}}+\left(K_{4}-\frac{K_{3}\tilde{A_3}}{\tilde{A_1}}\right)\frac{\Delta \mu Q}{T_{ent}}\right]~.
\end{equation} 

\noindent
From the thermodynamic relation of HSC in \cite{Karar:2017org}, it was found that the change in entanglement pressure($\Delta P_{x}$) does not appear in the expression of change in HSC, it is because of the presence of the $h_{xx}(r)+h_{yy}(r)$ term that eventually cancels out the $t_{d2}$ term, responsible for the pressure. Hence from eq.\eqref{mutual complexity} it is obvious that $\Delta P_{x}$ term will not be present in the expression for the change in mutual complexity. Our calculations clearly show the absence of $\Delta P_{x}$ term in the change of mutual complexity.
\section{Conclusion}\label{section 5}
In this paper we have studied various mixed state information theoretic measures for an excited state of Lifshitz spacetime in $3+1$-dimensions. The excited state of Lifshitz spacetime is obtained by applying constant perturbations (along boundary direction) on pure state of Lifshitz spacetime. The perturbation in the metric is given by the linear perturbation function (that is $h_{xx}(r)$ and $h_{yy}(r)$). It is to be noted that the form of the perturbation functions are different for different values of the dynamical exponent $z$. It is found that the perturbation functions have similar form for any arbitrary value of dynamical exponent except for $z=2$. This is why in this work we have computed our results for $z=2$ and any arbitrary value of $z$ separately. We have first holographically calculated the different mixed state information theoretic quantities for ground state Lifshitz spacetime, while doing the calculations we have choosen strip like subsystems in the boundary. The results of different information theoretic quantities computed for pure Lifshitz spacetime suggests that neither of the quantities exhibit any dynamical exponent dependence, this implies all the information theoretic quantities for pure Lifshitz spacetime has no $z$ dependence. Therefore, to understand how these quantities depend on the dynamical exponent, we have to consider excited Lifshitz spacetime over the pure one.
Then we have proceeded to compute the same quantities for excited Lifshitz spacetime. This computation shows how the different information theoretic quantities depend on the dynamical exponent. As mentioned earlier we have considered a strip like subsystem of length $l$ to compute all the mentioned quantities holographically. For both the cases (that is, for ground and as well as for the excited state), we have kept the subsystem length to be fixed. The underlying reason behind this lies in the fact that, we want to compare the results of different information theoretic quantities for the same setup. All the calculations we have done are first order in perturbation, that is first order in $h_{xx}(r)$ and $h_{yy}(r)$. We have already mentioned that to obtain the gravity dual of excited state of Lifshitz theory we have to apply a constant perturbation along the boundary direction. Due to this perturbation the turning point in the bulk associated to the subsystem length ($l$) changes as compared to the ground state. Therefore, we will denote the turning point for the excited state as $r_t =r^{(0)}_t +\delta r_t$, where $r^{(0)}_t$ is the turning point corresponding to the pure Lifshitz spacetime and $\delta r_t$ is the change in turning point due to perturbation in the geometry. As we are interested at the leading order change of different information theoretic quantities, we will ignore the change in turning point (that is, $\delta r_t$) arising due to the metric perturbation by setting $\delta r_t$ to be zero. Demanding the fact that $\delta r_t =0$, we have obtained a relation between the linearized perturbations (that is $h_{xx}(r)$ and $h_{yy}(r)$). This obtained relation is very useful to compute different results. As we have mentioned earlier that, we have ignored the change in the turning point corresponding to a subsystem of length $l$, this implies that the subsystem length in both the cases are kept to be same. Therefore, we have used the similar relation between the subsystem length and turning point of pure Lifshitz spacetime to compute different results of excited state. Here we would like to mention that the results of excited state are obtained for the same subsystem length as of the ground state Lifshitz spacetime. After obtaining the subsystem length in terms of turning point we have proceeded to compute the holographic entanglement entropy for a strip like subsystem. To compute the HEE we have followed the Ryu-Takayanagi prescription. As we are interested to the leading order correction in HEE due to the metric perturbation, we have used the relation between the perturbation function arises due to the fact that, we have ignored the change in turning point associated to the subsystem length due to perturbation. This makes our calculation simple. After obtaining a general form of HEE in terms of $h_{xx}(r)$ and $h_{yy}(r)$, we then compute this for different values of the dynamical exponent, namely, for $z=2$ and arbitrary value of $z$.\\ 
With this result of HEE in hand, we have computed the holographic mutual information (HMI) between two subsystems of equal length $l$ and which are separated by a distance $d$. We have computed the HMI for $z=2$ and also for an arbitrary value of $z$. The computation suggests that HMI between two subsystems is a divergent free quantity. Then we have computed EWCS by following the $E_P =E_W$ duality. This is also computed for different values of $z$, namely for $z=2$, and for arbitrary value of $z$. After computing these results of HMI and EWCS, we have graphically represented them to have a better understanding. Although the graphical representation of our results are bit tricky as the results contain arbitrary constants ($c_1$,$c_2$ and $t_{d2}$). To resolve this issue we have used the well known inequality $E_{W}(A:B)\geq \frac{I(A:B)}{2}$. This inequality will lead us to a bound satisfied by the constants for both the values of dynamical scaling exponent, that is, for $z=2$ and any arbitrary value of $z$. To obtain all our plots we have chosen the constants such that they will satisfy that bound. The plots suggests that the holographic mutual information vanishes for a particular value of the separation distance ($d_{c}$) for given choice of subsystem length. This fact is valid for $z=2$ and also for arbitrary value of $z$. We have also compared the results of EWCS and HMI between the excited state and ground state Lifshitz spacetime graphically. We have observed that, for both the values of dynamical exponent (for $z=2$ and for arbitrary value of $z$), the HMI for excited state vanishes earlier compared to the ground state Lifshitz spacetime. This implies the entanglement wedge associated with the excited state becomes disconnected earlier compared to the ground state. Then we have also compared the results of HMI and EWCS for different values of dynamical exponent (except for $z=2$). This plot suggests that the HMI vanishes earlier for higher values of dynamical exponent. Hence, the EWCS also shows the similar behaviour, that is, the entanglement wedge associated with the higher values of $z$ becomes disconnected earlier. \\
Then we have proceeded to compute another mixed state information theoretic quantity namely, entanglement negativity holographically. We have computed this quantity for two different set ups, namely, one for the adjacent subsystems and other for the disjoint subsystems. We have described that there are two different ways to compute the entanglement negativity holographically. In one of the proposals it was suggested that one can compute entanglement negativity by using the knowledge of entanglement wedge cross section. On the other hand, there is an another way to compute $E_N$ in terms of HEE. In our present work we have followed the second proposal to compute $E_N$. The results of $E_N$ (for different values of dynamical exponent) suggests that for adjacent subsystems entanglement negativity is always a divergent quantity. On the other hand for disjoint subsystems we have found that $E_N$ is always divergentless. Then we have plotted the results of $E_N$ for two disjoint subsystems of equal length $l$ and which are separated by a distance $d$. In these plots we have shown the variation of $E_N$ with respect to the distance of separation between the subsystems. We have plotted these results for two different sets of the dynamical exponent. To plot these results we have chosen the constants arises in the expressions of $E_N$, by following the similar procedure described in case of HMI and EWCS. The plot suggests that the entanglement negativity for two disjoint subsystems vanishes at a particular value subsystem separation ($d^{'}_c$) for fixed subsystem length and the dynamical exponent. This fact is valid for $z=2$ and as well as for arbitrary value of $z$. We would like to mention that, entanglement negativity vanishes for larger value of the separation distance compared to the HMI. This implies entanglement negativity measures the correlation between two subsystems even if they are not in the connected phase. We then proceed to compare the results of $E_N$ for excited state and ground state Lifshitz spacetime graphically. This plot suggests that for excited state (for any value of dynamical exponent) entanglement negativity vanishes earlier compared to the ground state. We have also compared the results of $E_N$ for different values of the dynamical exponent (except for $z=2$). This study reveals the fact that the critical separation between two subsystems decreases as we increase the value of the dynamical exponent. \\
Then we have moved on to compute the holographic sub region complexity of a subsystem of length $l$, by using the "Complexity=Volume" conjecture. With this result of HSC, we now proceed to compute mutual complexity for mixed states. Then we have plotted the result of mutual complexity between two subsystems with respect to the distance of separation between the subsystems. This plots suggests that for excited state, mutual complexity is always superadditive in nature, which implies $\Delta C$ is negative. We have also compared the results of mutual complexity between pure and excited state graphically. From the graph it is clear that the mutual complexity for excited state (for both the values of dynamical exponent) is always more negative compared to the ground state Lifshitz spacetime. Then we have proceeded to show the variation of mutual complexity with respect to the subsystem separation for different values of dynamical exponent. It is observed that the mutual complexity becomes more negative for higher values of dynamical exponent ($z$).\\ 
Using the stress tensor complex, we have then established first law like relations of entanglement thermodynamics for the change in EWCS, entanglement negativity and mutual complexity for both the values of dynamical scaling exponent, that is, for $z=2$ and any arbitrary value of $z$. We have seen that unlike the relativistic theories with conformal symmetries, the change in these quantities contains some extra terms like entanglement pressure and chemical potential. These extra terms in the thermodynamic relation appears due to the presence of massive gauge field in the theory. For both the values of dynamical scaling exponent, the thermodynamic relations only contain the change in $P_{x}$ term (normal to the entanglement region), however if we try to include total pressure ($P_{x}+P_{y}$) in our calculation it will not be possible to derive a first law like relation of entanglement thermodynamics. This is due to the fact that some of the parameters in the expression of $\Delta E_{W}$, $\Delta E_{N}$ and $\Delta C$ cannot be eliminated if we consider the total pressure. We have also noticed that the change in $E_{W}$, $E_{N}$ and $\Delta C$ are UV finite, otherwise thermodynamic relations would not exist. \\
It is important to recognize that instead of strip like subsystems, we may have done all calculations for a more compact entangling region. This will not alter the physical nature of the outcomes we have obtained. Any changes will only impact the coefficients of each term. We would also like to mention that all the analysis has been done for constant perturbations along the boundary direction. It will be very interesting to check what happens when someone considers general perturbations as mentioned in \cite{Ross:2009ar}. As a future direction, it would be interesting to study the Fidelity susceptibility and Fisher information for the excited state of Lifshitz spacetime. Furthermore, this study has gone someway towards describing mixed state entanglement measures in a holographic manner for Lifshitz systems which are non-relativistic analogue of conformal field theory.

\section*{Acknowledgments}
SP would like to thank SNBNCBS for Junior Research Fellowship, ARC thanks SNBNCBS for Senior Research Fellowship. AS would like to thank S.N. Bose National Centre for
Basic Sciences for the financial support through its Advanced Postdoctoral Research Programme.
\bibliographystyle{JHEP.bst}
\bibliography{citation_Lifshitz.bib}
\end{document}